\documentclass[reprint,aps,pre,amsmath,amssymb]{revtex4-1}
\usepackage{graphicx}

\usepackage[caption=false]{subfig}
\usepackage{hyperref}
\usepackage{bm}
\usepackage{color}

\renewcommand{\vec}[1]{\bm{#1}}

\begin{document}

\title{Persistent exclusion processes: inertia, drift, mixing and correlation}
\author{Stephen Zhang} \author{Aaron Chong} \author{Barry D. Hughes} \email{barrydh@unimelb.edu.au}
\affiliation{School of Mathematics and Statistics, University of Melbourne, Victoria 3010, Australia}
\date{\today}

\begin{abstract}
In many biological systems, motile agents exhibit random motion with short-term directional persistence, together with crowding effects arising from spatial exclusion. We formulate and study a class of lattice-based models for multiple walkers with motion persistence and spatial exclusion in one and two dimensions, and use a mean-field approximation to investigate relevant population-level partial differential equations in the continuum limit. We show that this model of a persistent exclusion process is in general well described by a nonlinear diffusion equation. With reference to results presented in the current literature, our results reveal that the nonlinearity arises from the combination of motion persistence and volume exclusion, with linearity in terms of the canonical diffusion or heat equation being recovered in either the case of persistence without spatial exclusion, or spatial exclusion without persistence. We generalise our results to include systems of multiple species of interacting, motion-persistent walkers, as well as to incorporate a global drift in addition to persistence. These models are shown to be governed approximately by systems of nonlinear advection-diffusion equations. By comparing the prediction of the mean-field approximation to stochastic simulation results, we assess the performance of our results. Finally, we also address the problem of inferring the presence of persistence from simulation results, with a view to application to experimental cell-imaging data. 
\end{abstract}

\maketitle

\section{Introduction}

Random walks are a broad class of stochastic processes relevant for modelling a wealth of phenomena \cite{Hughes1995, Weiss1994}. An area of application which has received significant interest is the use of such models for the study of cellular motility \cite{Codling2008, Mort2016, Krummel2016, Baker2014}. Much of the existing literature on random walks considers a single walker. However, in some contexts, it is useful to model a population of walkers on a lattice, for which it is necessary to account for interactions \cite{Simpson2009, Codling2008}. As a first model of a system of interacting walkers (sometimes referred to as agents), occupancy of a given lattice site by multiple agents is forbidden, with attempted moves by agents into already occupied sites being aborted. If the process evolves in continuous time with each individual agent waiting an exponentially distributed time between move attempts, any two agents will attempt simultaneous moves with probability zero and so clashing move attempts do not occur. Alternatively, we may allow the process to evolve in discrete time with a random sequential update procedure: if there are $N$ agents present, we make $N$ sequential random choices of agent with replacement, offering each chosen agent an opportunity to move. Once $N$ such offers have been made, we increment time by a constant amount. When a simple random walk stepping rule for individual agents is combined with either of these two time evolution protocols and the abortion of moves onto occupied sites, we obtain a model called the simple exclusion process (SEP). A number of rigorous results are available for this particular model \cite{Liggett1985, Liggett1999, Komorowski2007}, the most important of these being that in an appropriate continuum limit, the probability of occupancy of a location evolves under the classical linear diffusion (or heat) equation \cite{Liggett1999}.

For most systems of interacting agents apart from the SEP, exact continuum limit partial differential equations (PDEs) have been not been established and indeed may not exist due to subtle effects of inter-agent correlation (which happen to conveniently remove themselves from some questions that can be answered for the SEP). However, useful approximate continuum limit PDEs can be obtained by mean-field arguments in which correlation effects are either neglected (see, e.g.,  \cite{Simpson2009}, although the literature in this area is quite extensive), or modelled in a simple approximate manner \cite{Markham2013}. For the SEP, the simplest mean-field approach actually produces the correct PDE \cite{Simpson2009, Simpson2009b}. Further results have been developed for variations on the simple exclusion process, including incorporation of additional bias effects and accounting for interactions between multiple species of walkers \cite{Simpson2009b}, as well as consideration of alternative interaction rules \cite{Penington2011, Almet2015, Nan2018, Fernando2010}. In many cases the predictions of the mean-field PDE agree well with simulation averages, especially at low densities of agents, but the quality of agreement degrades when inter-agent interactions have a significant attractive component \cite{Fernando2010}, or when agents reproduce at random as well as moving at random.

In many scenarios including modelling of vehicle movement \cite{Michelini2008}, chemical reaction kinetics \cite{Vilensky1994}, cell motility \cite{Baker2014} and animal movement \cite{Wu2000}, it is clear that agents do not move completely at random but exhibit short-term persistence in the direction of motion, thus giving rise to the notion of a persistent random walk (PRW) \cite{Codling2008, Gorelik2014, Wu2000}. The case of non-interacting walkers has received substantial attention \cite{Patlak1953, Masoliver2017,Codling2008, Renshaw1981, Henderson1984}. In particular, it is known that the non-interacting persistent random walk is governed by a linear diffusion equation \cite{Renshaw1981, Henderson1984}. However, current work considering multiple interacting persistent random walks remains sparse \cite{Gavagnin2018, Teomy2019, Teomy2019b}. 

In the persistent exclusion process, we consider swarms of interacting, motion-persistent agents. Mean-field approximations may be used to derive an approximate continuum description of such a system at the macroscopic level. Gavagnin and Yates \cite{Gavagnin2018} presented such an analysis for one model formulation in which individual agents follow an explicit velocity-jump process. Agents possess one of four possible polarisations on a two-dimensional lattice, and either move following a polarisation-dependent rule for step directions, or repolarise uniformly. Movement and repolarisation are driven by independent Poisson processes, and in the continuum limit this model yields a system of four nonlinear PDEs describing the dynamics of agents in each of the four possible orientations. Even more recently, Teomy and Metzler \cite{Teomy2019, Teomy2019b} reported results for transport in a general model of exclusion processes in which agents possess a finite memory of \emph{attempted} moves and move in a memory-dependent manner. The authors showed that in the continuum limit such a system may be described by nonlinear advection-diffusion equations, verifying their results for the case of steady-state transport in one dimension. 

In this paper, we present and analyse in one and two dimensions a related but distinct model for a persistent exclusion process where the motion persistence of individual agents is incorporated as a single-step memory of the last \emph{successful} move taken \cite{Renshaw1981, Patlak1953, Jones2015, Galanti2013, Hughes1995, Weiss1994}, and hence the sequence of successful moves made by each agent follows a strict Markov rule. We begin by formulating our agent-based model for a number of cases, from which we then derive population-level descriptions in the continuum limit. For the basic case of a single species of motion-persistent agents, we find that such systems are approximately governed by nonlinear diffusion equations. Although our analysis yields results resembling those presented by Teomy and Metzler \cite{Teomy2019}, key differences are revealed regarding the descriptions of population-level behaviour. We extend our analysis to consider generalisations of the basic model involving multiple species of interacting agents as well as a superimposed global drift effect, showing that in general such systems are approximately governed by systems of nonlinear advection-diffusion equations. By comparing these continuum results to simulation data, we verify that our derived continuum approximations hold for evolving, inhomogeneous systems. Finally, we investigate methods for inferring the presence of persistence from simulation results, with a view to application to experimental cell-imaging data \cite{Mort2016,Agnew2014}.

\section{Agent-based model}\label{sec:abm}

\subsection{One-dimensional model}

We consider $N \gg 1$ agents on $\mathbb{Z}$, the usual one-dimensional lattice of sites with integer coordinates. Agents on the lattice are able to move right (R) or left (L), and to each agent we ascribe the direction of its last successful move to be its \emph{orientation}. We thus partition the agent population by orientation into two groups denoted R and L. At each timestep, we sample $N$ agents with replacement following the random sequential update procedure \cite{Simpson2009} previously described. Each sampled agent, with probability $P \in (0, 1]$, will attempt to move. If the destination site is vacant the attempted move is successful, otherwise it is aborted. Move attempts are made with the following assigned probabilities:
\emph{in} the direction of orientation with probability $(1+\varphi)/2$, and 
\emph{against} the direction of orientation with probability $(1 - \varphi)/2$.
We illustrate this in Fig.~\ref{fig:movementrule}(a) for a right-oriented agent. The parameter $\varphi \in [-1, 1]$ controls the extent of motion persistence. We make note that the case $\varphi < 0$ corresponds to a `negative' persistence effect where agents have a propensity to undo their previous move. 

\subsection{Two-dimensional model}\label{sec:twodim}

\begin{figure}
\begin{center}
\subfloat[]{\includegraphics[width = 0.45\linewidth]{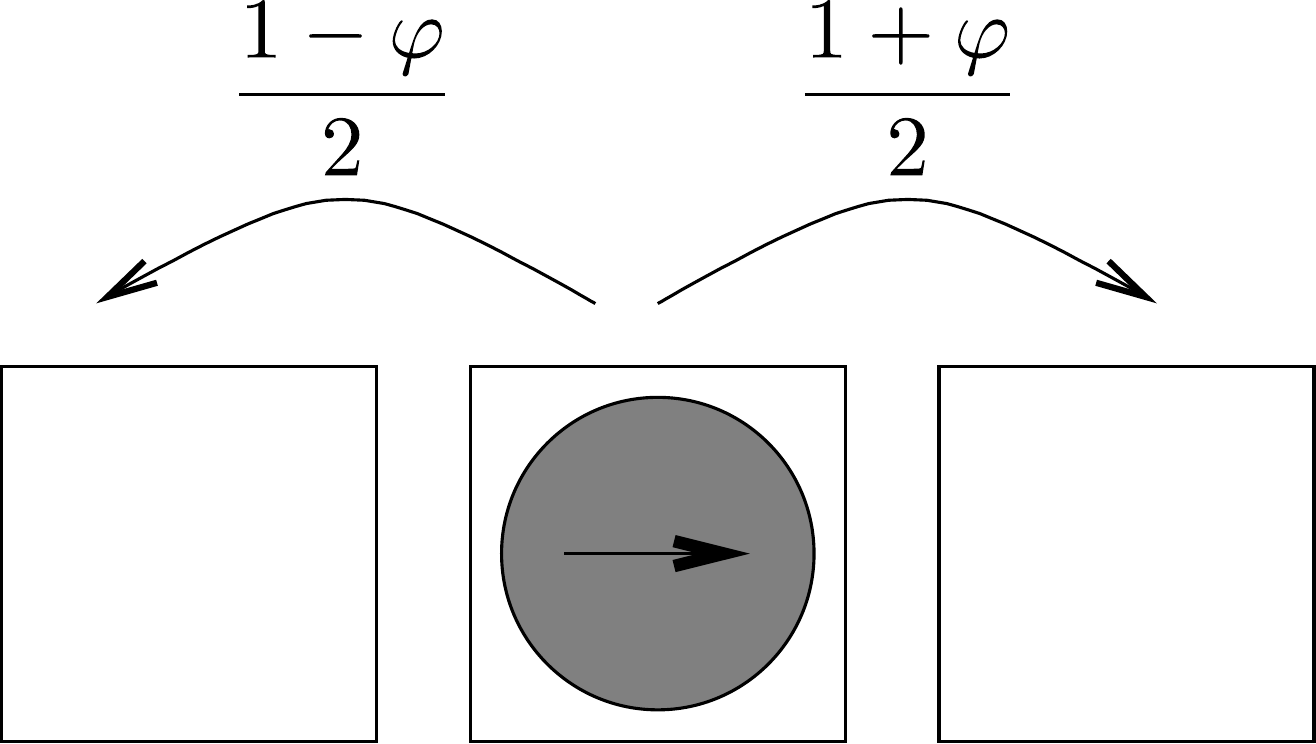}}\\ 
\subfloat[]{\includegraphics[width = 0.45\linewidth]{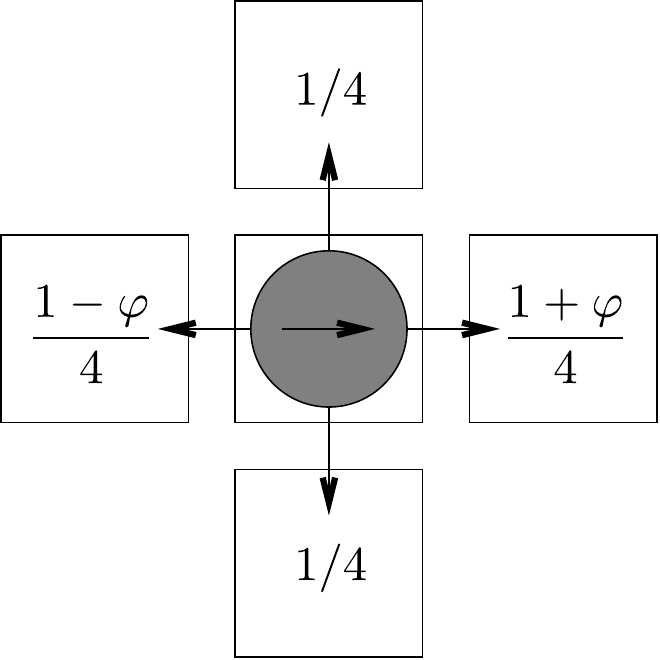}}
\end{center}
\caption{Movement attempt probabilities for a rightwards (R) oriented agent for (a) one-dimensional and (b) two-dimensional agent-based models, where parameter $\varphi \in [-1, 1]$ controls the extent of motion persistence}
\label{fig:movementrule}
\end{figure}

Consider now the square lattice $\mathbb{Z}^2$. We may extend the one-dimensional model to now consider four possible orientations, right (R), left (L), up (U) and down (D). An attempted move is taken \emph{in} the direction of orientation with probability $(1 + \varphi)/4$, \emph{against} the direction of orientation with probability $(1 - \varphi)/4$, and in one of the two choices \emph{orthogonal} to the direction of orientation each with probability $1/4$.
We illustrate this in Fig.~\ref{fig:movementrule}(b) for an agent whose most recent move was to the right.

Note that in this two-dimensional case, the parameter $\varphi$ controls only the propensity of agents to move in the direction parallel to their orientation, and that we enforce that agents move orthogonal to their orientation with a fixed probability. We have obtained corresponding results for a more general two-parameter model from which the simpler model can be derived (see Appendix \ref{app:generalised_model}), but for simplicity we limit our discussion to the single-parameter model. 

\subsection{Two-dimensional model with global drift}\label{sec:globaldrift}

\begin{figure}
    \centering
    \includegraphics[width = \linewidth]{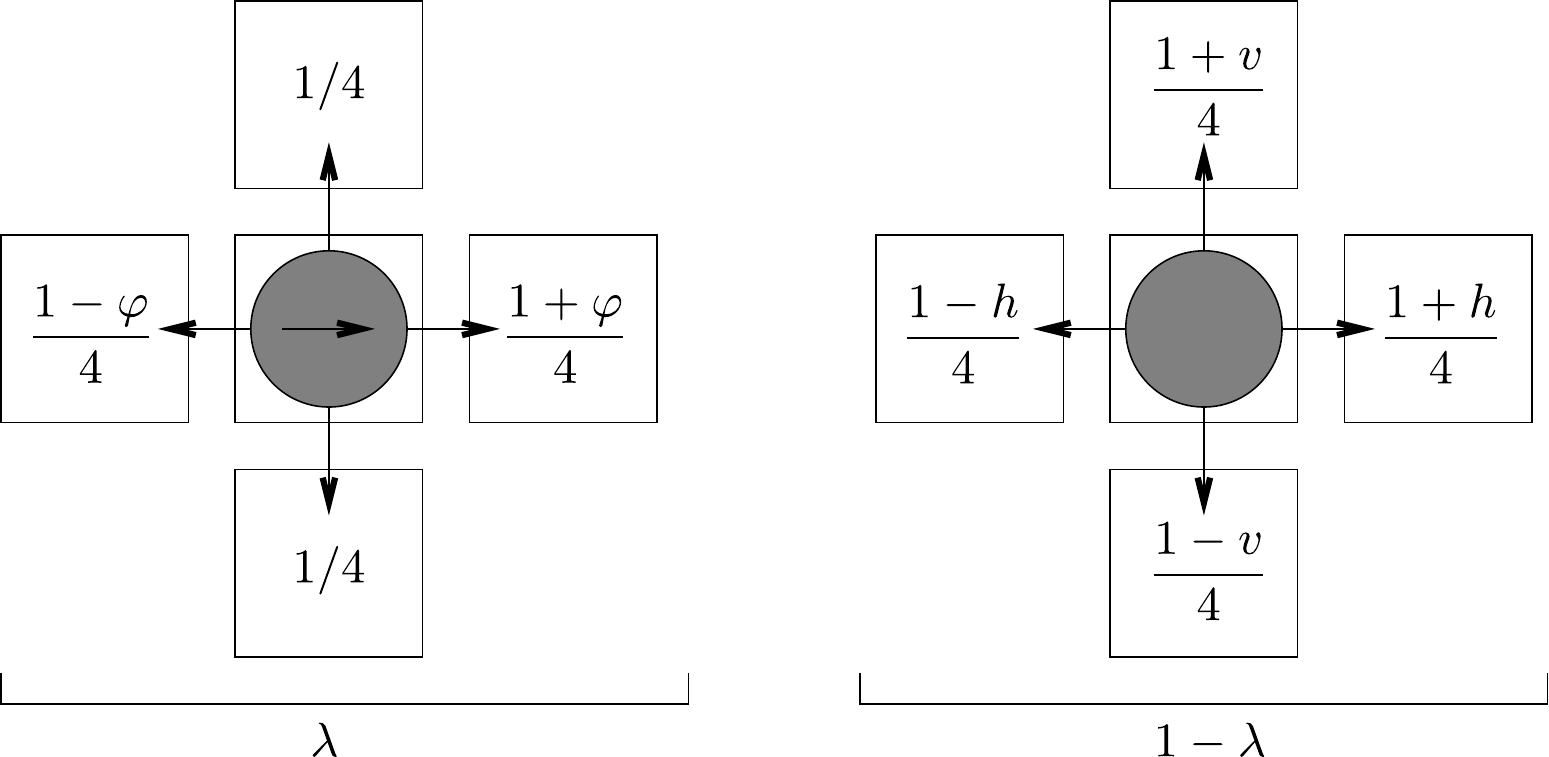}
    \caption{Movement attempt probabilities for a rightwards (R) oriented agent for the two-dimensional agent-based model with global drift. Parameter $\varphi \in [-1, 1]$ controls the extent of motion persistence, $\lambda \in [0, 1]$ controls the relative contributions of local persistence and global drift, and $h, v \in [-1, 1]$ control the direction and magnitude of global drift.}
    \label{fig:two-parameter-movement-rule}
\end{figure}

We generalise the two-dimensional model to incorporate both a local persistence and global drift. We introduce an additional parameter $\lambda \in [0, 1]$ that specifies the relative contribution of local persistence and global drift, and parameters $h, v \in [-1, 1]$ controlling the direction and magnitude of global drift in the horizontal and vertical directions respectively. An agent offered the opportunity to move chooses its move as follows. With probability $\lambda$, choose how to move as in Section \ref{sec:twodim}.
Alternatively with probability $1 - \lambda$, attempt a move that respects the imposed global drift: \emph{right} with probability \mbox{$(1 + h)/4$},  \emph{left} with probability \mbox{$(1-h)/4$}, \emph{up} with probability \mbox{$(1+v)/4$}, or \emph{down} with probability \mbox{$(1 - v)/4$}. This movement rule is illustrated in Fig.~\ref{fig:two-parameter-movement-rule} for an agent whose most recent move was to the right.

It is important to note here that in our agent-based model, agents only remember \emph{successful} moves, with aborted move attempts being `forgotten'. This is in contrast with the model presented by Teomy and Metzler \cite{Teomy2019}, in which the memory is updated regardless of the outcome of a move attempt. The resulting differences for bulk system behaviour are immediately appreciable. For instance, in one dimension with extreme negative persistence ($\varphi = -1$), Teomy and Metzler report that mean square displacement (MSD) exhibits $\sqrt{t}$ behaviour for starting densities above $1/2$. For our model, however, such behaviour is never possible in the equivalent setting. 

\section{Continuum-limit analysis}
\subsection{Single species}\label{sec:singlespecies}
For brevity, we show the derivation of the continuum-limit PDE in the two-dimensional case, as the analysis for the one-dimensional model is analogous. We model the agent-based model as a random walk with internal states \cite{Hughes1995, Weiss1994}. As mentioned earlier, we partition agents by orientation into subpopulations henceforth labelled R, L, U and D. We introduce $\vec{P}_n(i, j)$ as the vector of occupancy probabilities for the lattice site $(i, j)$ at time step $n$, with
\begin{equation}
	\vec{P}_n(i, j) = \begin{bmatrix}R_n(i, j) & L_n(i, j) & U_n(i, j) & D_n(i, j)\end{bmatrix}^\top, 
\end{equation}
where $R_n, L_n, U_n$ and $D_n$ are the respective occupancy probabilities for right, left, up and down-oriented agents. We also define the total occupancy $C_n(i, j)$ to be
\begin{equation}
    C_n(i, j) = R_n(i, j) + L_n(i, j) + U_n(i, j) + D_n(i, j).
\end{equation}

Under a mean-field approximation, we assume that average lattice occupancies of distinct sites are independent, and thus the evolution of $\vec{P}_n$ with time across the lattice is approximately governed by the equation
\begin{equation}
	\vec{P}_{n+1}(\vec{k}) = \vec{T}_n(\vec{k} | \vec{k}) \vec{P}_n(\vec{k}) \enskip + \sum_{\vec{k'} \in \mathcal{N}(\vec{k})} \vec{T}_n(\vec{k} | \vec{k'}) \vec{P}_n(\vec{k'}), 
	\label{eqn:master}
\end{equation}
where $\mathcal{N}(\vec{k}) = \left\{ \vec{k} + \vec{e_1}, \vec{k} - \vec{e_1}, \vec{k} + \vec{e_2}, \vec{k} - \vec{e_2}\right\}$ denotes the set of sites immediately adjacent to $\vec{k}$, and $\vec{e_{1, 2}}$ denote the basis vectors in the $x$ and $y$ directions respectively. $\vec{T}_n(\vec{k} | \vec{k'})$ is a matrix of transition probabilities for the transition from site $\vec{k'}$ to $\vec{k}$, which is in general not constant but dependent on $\vec{k}, \vec{k'}$ and $n$ as well as $\vec{P}_n$. From this, we obtain the master equation in Table \ref{table_of_long_formulae}(a).
\begin{turnpage}
\begin{table*}
(a) Master equation in discrete lattice coordinates 
\begin{equation*}
	\begin{alignedat}{2}
	\vec{P}_{n+1}(i, j) &= (1-P)\vec{P}_n(i, j)
	+ \dfrac{P}{4}\left\{
		\mathrm{diag}\begin{bmatrix} 1+\varphi \\ 1-\varphi \\ 1 \\ 1 \end{bmatrix}C_n(i+1, j) +
		\mathrm{diag}\begin{bmatrix} 1-\varphi \\ 1+\varphi \\ 1 \\ 1 \end{bmatrix}C_n(i-1, j)
			+ \mathrm{diag}\begin{bmatrix} 1 \\ 1 \\ 1+\varphi \\ 1-\varphi \end{bmatrix}C_n(i, j+1) + 
			\mathrm{diag}\begin{bmatrix} 1 \\ 1 \\ 1-\varphi \\ 1+\varphi \end{bmatrix}C_n(i, j-1)\right\}\vec{P}_n(i, j) \\
		&+ \dfrac{P[1-C_n(i, j)]}{4} \left\{ 
			\begin{bmatrix} 1+\varphi & 1-\varphi & 1 & 1 \\ 0&0&0&0 \\ 0&0&0&0 \\ 0&0&0&0 \end{bmatrix} \vec{P}_n(i-1, j) 
			+ \begin{bmatrix} 0&0&0&0 \\ 1-\varphi & 1+\varphi & 1 & 1 \\ 0&0&0&0 \\ 0&0&0&0 \end{bmatrix} \vec{P}_n(i+1, j) 
			+
			\begin{bmatrix} 0&0&0&0 \\ 0&0&0&0 \\ 1 & 1 & 1+\varphi & 1-\varphi \\ 0&0&0&0 \end{bmatrix} \vec{P}_n(i, j-1) 
				+ \begin{bmatrix} 0&0&0&0 \\ 0&0&0&0 \\ 0&0&0&0 \\ 1 & 1 & 1-\varphi & 1+\varphi \end{bmatrix} \vec{P}_n(i, j+1)	
		\right\}
	\end{alignedat}
	\label{eqn:master_1param}
\end{equation*}

(b) Master equation in continuous coordinates in the original basis
\begin{equation*}
	\begin{alignedat}{2}
\tau\dfrac{\partial \vec{P}}{\partial t} 	&=\dfrac{P}{4}\left\{ 2\varphi \Delta \left(
		\mathrm{diag} \begin{bmatrix} 1 \\ -1 \\ 0 \\ 0 \end{bmatrix} \dfrac{\partial C}{\partial x}(x, y, t) + 
		\mathrm{diag} \begin{bmatrix} 0 \\ 0 \\ 1 \\ -1 \end{bmatrix} \dfrac{\partial C}{\partial y}(x, y, t) \right) + \Delta^2 \left(\nabla^2 C(x, y, t)\right) \vec{I} \right\}\vec{P}(x, y, t)\\
		& + \dfrac{P[1-C(x, y, t)]}{4} \left\{
		\begin{bmatrix}-3+\varphi & 1-\varphi & 1 & 1 \\ 1-\varphi & -3+\varphi & 1 & 1 \\ 1 & 1 & -3+\varphi & 1-\varphi \\ 1 & 1 & 1-\varphi & -3+\varphi \end{bmatrix} \vec{P}(x, y, t) 
		 + \Delta\left(\begin{bmatrix}-1-\varphi & -1+\varphi & -1 & -1 \\ 1-\varphi & 1+\varphi & 1 & 1 \\  0&0&0&0 \\  0&0&0&0 \end{bmatrix} \dfrac{\partial \vec{P}}{\partial x} + 
		\begin{bmatrix}  0&0&0&0 \\  0&0&0&0 \\ -1 & -1 & -1-\varphi & -1+\varphi \\ 1 & 1 & 1-\varphi & 1+\varphi \end{bmatrix} \dfrac{\partial \vec{P}}{\partial y} \right) \right. \\
		& \left.\hspace{2.35cm} +  \dfrac{\Delta^2}{2}\left( \begin{bmatrix} 1+\varphi & 1-\varphi & 1 & 1 \\ 1-\varphi & 1+\varphi & 1 & 1 \\ 0&0&0&0 \\  0&0&0&0\end{bmatrix} \dfrac{\partial^2 \vec{P}}{\partial x^2} + 
		\begin{bmatrix}  0&0&0&0 \\  0&0&0&0 \\ 1 & 1 & 1+\varphi & 1-\varphi \\ 1 & 1 & 1-\varphi & 1+\varphi \end{bmatrix} \dfrac{\partial^2 \vec{P}}{\partial y^2}\right)\right\}		
	\end{alignedat}
	\label{app:continuous}
\end{equation*}
(c) Master equation in continuous coordinates in the  eigenvector basis 
\begin{align*}
	\begin{split}
		\tau \dfrac{\partial \vec{\Phi}}{\partial t} &=
			\dfrac{P}{4} \left\{ 2\varphi \Delta \left( 
				\begin{bmatrix} 0& 0& 1 &0 \\ 0& 0& 1 &0 \\ 1/2 & 1/2 &0 &0 \\ 0&0&0&0 \end{bmatrix} \dfrac{\partial C}{\partial x} 
			+	\begin{bmatrix} 0& 0& 0& 1 \\ 0& 0& 0& -1 \\ 0&0&0&0 \\ 1/2 & -1/2 &0&0\end{bmatrix} \dfrac{\partial C}{\partial y}
			\right)	
			+	\Delta^2\left(\nabla^2 C\right) \vec{I}
			\right\}\vec{\Phi}(x, y, t) \\
			& + \dfrac{P[1 - C(x, y, t)]}{4} \left\{ \mathrm{diag}\begin{bmatrix} 0 \\ -4 \\ -4 + 2\varphi \\ -4 + 2\varphi \end{bmatrix} \vec{\Phi}(x, y, t) 
			+ \Delta\left( 
				\begin{bmatrix} 0& 0& -2\varphi &0 \\ 0& 0& -2\varphi &0 \\ -2 &0 &0 &0 \\ 0&0&0&0 \end{bmatrix} \dfrac{\partial \vec{\Phi}}{\partial x} 
			+ 	\begin{bmatrix} 0& 0& 0& -2\varphi \\ 0& 0& 0& 2\varphi \\ 0&0&0&0 \\ -2 &0&0&0 \end{bmatrix} \dfrac{\partial \vec{\Phi}}{\partial y} 
			\right) 
			+ \dfrac{\Delta^2}{2} \left(
				\begin{bmatrix} 2 &0&0&0 \\ 2 &0&0&0 \\ 0&0& 2\varphi &0 \\ 0&0&0&0 \end{bmatrix} \dfrac{\partial^2 \vec{\Phi}}{\partial x^2} 
			+ 	\begin{bmatrix} 2 &0&0&0 \\ -2 &0&0&0 \\ 0&0&0&0 \\ 0&0&0& 2\varphi \end{bmatrix} \dfrac{\partial^2 \vec{\Phi}}{\partial y^2}
			\right)\right\}
	\end{split}
	\label{app:continuous_transformed}
\end{align*}
\caption{Master equations (a) in discrete lattice coordinates (b) in continuous coordinates in the original basis (the equation for $\vec{P}$) and (c) the eigenvector basis (the equation for $\vec{\Phi}$). In (b) and (c), terms which are $\mathcal{O}(\tau^2)$  or $\mathcal{O}(\Delta^3)$ are negligible in the continuum limit compared to terms exhibited here and are not shown. Here $\vec{P}$ and $\vec{\Phi}$ and their partial derivatives are all evaluated at $(x,y,t)$.}\label{table_of_long_formulae}
\end{table*}
\end{turnpage}

Following the typical procedure \cite{Renshaw1981, Codling2008, Simpson2009, Simpson2009b, Gavagnin2018}, we introduce continuous coordinates $(x, y, t)$, a lattice spacing $\Delta$, and time step $\tau$. We take $x = \Delta i, y=  \Delta j, t = \tau n$. By making a formal substitution of variables, our vector of occupancy probabilities becomes
\begin{equation}
	\vec{P}(x, y, t) = \begin{bmatrix}R(x, y, t) & L(x, y, t) & U(x, y, t) & D(x, y, t)\end{bmatrix}^\top,
\end{equation}
and the total occupancy becomes 
\begin{equation}
	C(x, y, t) = R(x, y, t) + L(x, y, t) + U(x, y, t) + D(x, y, t).
\end{equation}
We then substitute formally into Table \ref{table_of_long_formulae}(a) and take Taylor expansions of $\vec{P}$ and $C$ to first order in $t$ and second order in $(x, y)$. Cancelling a term in $\vec{P}(x,y,t)$ from both sides and rearranging, we arrive at the system 
in continuous variables shown in Table \ref{table_of_long_formulae}(b). 
\newpage
We now perform a judicious change of basis for the system into the eigenvectors of the matrix
\begin{equation}
\begin{bmatrix}-3+\varphi & 1-\varphi & 1 & 1 \\ 1-\varphi & -3+\varphi & 1 & 1 \\ 1 & 1 & -3+\varphi & 1-\varphi \\ 1 & 1 & 1-\varphi & -3+\varphi \end{bmatrix}
\end{equation}
found to feature prominently within the continuous-variable master equation in Table \ref{table_of_long_formulae}(b). We use the change of basis matrix 
\begin{align}
	\vec{M}_{\mathrm P\to\Phi} = \begin{bmatrix} 1 & 1 & 1 & 1 \\ 1 & 1 & -1 & -1 \\ 1 & -1 & 0 & 0 \\ 0 & 0 & 1 & -1 \end{bmatrix}, 
		\label{eqn:transformation_matrix}
\end{align}
to form from $\vec{P}$ the transformed quantity $\vec{\Phi}$ given by 
\begin{equation}
	\vec{\Phi} = \vec{M}_{\mathrm P\to\Phi} \vec{P} = \begin{bmatrix} R + L + U + D \\ (R + L) - (U+D) \\ R-L \\ U-D \end{bmatrix}. 
		\label{eqn:transformed_def}
\end{equation}
Note that all components of $\vec{\Phi}$ except the first correspond to differences between subpopulation occupancy probabilities. By symmetry of the problem, we reason that as $\Delta \to 0$ in the continuum limit, these components should vanish since no individual direction is preferred above others. We thus formulate an ansatz for the form of $\vec{\Phi}$ in which the vanishing components are scaled by $\Delta$, that is, 
\begin{equation}
	\vec{\Phi} = \begin{bmatrix} \phi_1 && \Delta \phi_2 && \Delta \phi_3 && \Delta \phi_4\end{bmatrix}^\top,
	\label{eqn:scaled_phi_def}
\end{equation}
where $\phi_i = \phi_i(x, y, t)$ for $1 \leq i \leq 4$.
From this, we obtain 
\begin{equation}
	\vec{P} = \dfrac{\phi_1}{4} \begin{bmatrix}1\\1\\1\\1 \end{bmatrix} + 
				\dfrac{\Delta \phi_2}{4} \begin{bmatrix} 1 \\ 1 \\ -1 \\ -1 \end{bmatrix} +
					\dfrac{\Delta \phi_3}{2} \begin{bmatrix} 1 \\ -1 \\ 0 \\ 0 \end{bmatrix} + 
						\dfrac{\Delta \phi_4}{2} \begin{bmatrix} 0 \\ 0 \\ 1 \\ -1 \end{bmatrix}.
\end{equation}

Transforming the system in Table \ref{table_of_long_formulae}(b), we obtain the simplified system in Table \ref{table_of_long_formulae}(c). We first apply the continuum limit $\Delta, \tau \to 0$ for components $\phi_2, \phi_3, \phi_4$, which respectively yields
\begin{align}
	(1 - C)P\phi_2(x, y, t) &= 0, 
	\label{eqn:vanishing_component_eqn_1}\\
	\dfrac{P}{4}\left\{ \dfrac{\partial C}{\partial x} \varphi \phi_1 - 2(1 - C)\left[(2 - \varphi)\phi_3 + \dfrac{\partial \phi_1}{\partial x}\right] \right\} &= 0, 
	\label{eqn:vanishing_component_eqn_2}\\
	\dfrac{P}{4}\left\{ \dfrac{\partial C}{\partial y} \varphi \phi_1 - 2(1 - C)\left[ (2 - \varphi)\phi_4 + \dfrac{\partial \phi_1}{\partial y} \right]\right\} &= 0.
	\label{eqn:vanishing_component_eqn_3}
\end{align}
Further, the equation for the first component $\phi_1$ (without taking any limits) transpires to be
\begin{widetext}
\begin{align}
\tau\frac{\partial\phi_1}{\partial t} +\mathcal{O}(\tau^2) =
		\dfrac{P\Delta^2}{4} \left\{ \nabla^2 C \phi_1 + 2\varphi \vec{\nabla} C \cdot \begin{bmatrix} \phi_3 \\ \phi_4 \end{bmatrix} 
		+ (1 - C) \left[ -2\varphi \left( \dfrac{\partial \phi_4}{\partial y} + \dfrac{\partial \phi_3}{\partial x}\right) +\nabla^2 \phi_1\right]  \right\}+ \mathcal{O}(\Delta^3).
	\label{eqn:firstcomponent}
\end{align}
\end{widetext}

\noindent Solving (\ref{eqn:vanishing_component_eqn_1}), (\ref{eqn:vanishing_component_eqn_2}), (\ref{eqn:vanishing_component_eqn_3}) for $\phi_2, \phi_3, \phi_4$, we differentiate to obtain $\partial_y \phi_4$ and $\partial_x \phi_3$. Substituting into (\ref{eqn:firstcomponent}), we take the continuum limit in the usual manner holding $\Delta^2/\tau$ constant \cite{Codling2008, Simpson2009, Simpson2009b, Renshaw1981, Gavagnin2018}. Noting that $C = \phi_1$, we obtain a nonlinear diffusion equation of the general form
\begin{equation}
\dfrac{\partial u}{\partial t} = \vec{\nabla} \cdot \left[\mathcal{D}(u)\vec{\nabla} u\right],\label{eqn:pde_1param_full_eqn}
\end{equation}
where we take $u = \phi_1$ and the nonlinear diffusivity is given by
\begin{align}
\mathcal{D}(u) &= D \left( \dfrac{2 + \varphi}{2 - \varphi} \right) (1 - \varphi u ), &D &= \lim_{\Delta, \tau \to 0} \dfrac{\Delta^2 P}{4\tau}.\label{eqn:diffusivity_single_species}
\end{align}
In the absence of persistence, we set $\varphi = 0$ and our diffusivity reduces to $\mathcal{D}(u) = D$, that is, we recover the linear diffusion equation as a description of the simple exclusion process \cite{Simpson2009, Simpson2009b}.

In the case of the one-dimensional model, similar arguments also produce the nonlinear diffusion equation given in (\ref{eqn:pde_1param_full_eqn}) where we interpret $\vec{\nabla}$ as acting in one dimension, and  
\begin{align}
 \mathcal{D}(u) &= D\left(\dfrac{1 + \varphi}{1-\varphi}\right) (1-2\varphi u),&D &= \lim_{\Delta, \tau \to 0} \dfrac{\Delta^2 P}{2\tau}. \label{eqn:pde_1d_1param_full_diffusivity}
\end{align}

We note in both cases that for $\varphi > 0$ ($\varphi < 0$), the diffusivity $\mathcal{D}$ is linear in the density $u$ and decreases (increases) with increasing density, whereas for $\varphi = 0$ in the non-persistent case, the diffusivity remains density-independent. Holding $u$ constant, the diffusivity is also an increasing function of the persistence parameter $\varphi$. Introduction of motion-persistence has thus led to the nontrivial result of a diffusivity intriguingly dependent on both the density $u$ and extent of persistence $\varphi$. 

In the one-dimensional case, the diffusivity becomes negative for $u > (2\varphi)^{-1}$. Together with the observation that on a one-dimensional lattice the relative ordering of agents must remain fixed, we conclude that at large $\varphi$ and $u$, spatial correlation of site occupancies on the one-dimensional lattice become significant and so mean-field arguments employed in this analysis are invalidated. We also observe that $\mathcal{D}$ becomes unbounded as $\varphi \to 1$. This corresponds qualitatively to a degenerate case for the one-dimensional discrete process in which agents are unable to change orientation and so are forced to move in a fixed direction. Noting that non-interacting walkers subject to this movement rule would exhibit ballistic rather than diffusive motion, we suggest that this is in some way reflected in the blowing up of $\mathcal{D}$. Thus, we expect the continuum results in one dimension to be useful mostly for relatively dilute systems, or for low persistence. For the two-dimensional case the diffusivity is always non-negative.

Whilst our analysis yields a diffusivity linear in the density $u$ for both one and two dimensions, the analogous result for diffusivity in one dimension obtained by Teomy and Metzler \cite{Teomy2019} varies quadratically with density. In particular, the authors reported a critical density ($1/2$ in the mean-field approximation), below which the diffusivity increases with increasing persistence, and above which the diffusivity instead decreases with increasing persistence. In contrast, our model displays no such behaviour, suggesting that the reported behaviour is characteristic of the specific model formulation used where agents keep track of the last \emph{attempted} move.

The model presented by Gavagnin and Yates \cite{Gavagnin2018} yielded systems of four coupled PDEs which required explicit solutions for each subpopulation in the continuum limit. On the other hand, analysis of memory-based systems yields only a single equation in the total population density, e.g. (\ref{eqn:pde_1param_full_eqn}) which naturally lends itself to interpretation as a nonlinear diffusion equation. This is a result of the scaling argument (\ref{eqn:scaled_phi_def}), where we take the limit in which the occupancy probability at each position is uniformly distributed among the possible orientations. In Section \ref{sec:results}, we find that the match between the resulting continuum description and simulation is very good. Nonetheless, we hypothesise that in principle it may be possible to derive a system of PDEs describing the individual subpopulations and thus account for cases such as initial conditions with non-uniform distribution of agent orientation. From simulations of the agent-based model however, we find that for mild persistence such initial anisotropies generally dissipate quickly to recover the isotropic case considered in our continuum-limit analysis. 

Curiously, it is known that both the non-interacting persistent random walk \cite{Renshaw1981, Henderson1984} and the non-persistent exclusion process \cite{Simpson2009} can be described by a canonical linear diffusion equation in the continuum limit. Based on this knowledge, our findings indicate that only upon combining both persistence and exclusion does the system yield behaviour governed by a \emph{nonlinear} diffusion equation.  

\subsection{Multiple species}\label{sec:multispecies}

Consider now $M$ distinct species of random walkers on the two-dimensional lattice, each of which possess species-specific parameter values for motility $P^{(k)}$ and persistence $\varphi^{(k)}$, $1 \leq k \leq M$. Subject to these parameters, agents of each species are governed by the movement rules described in Section \ref{sec:abm}.

For the $k$th species, we introduce the vector of occupancy probabilities
\begin{equation}
	\vec{P}^{(k)}_n(i, j) = \begin{bmatrix} R^{(k)}_n(i, j) & L^{(k)}_n(i, j) & U^{(k)}_n(i, j) & D^{(k)}_n(i, j)\end{bmatrix}^\top,
\end{equation}
where $R^{(k)}_n(i, j)$ is the occupancy probability for right-oriented agents of species $k$, and so on. Further, we define the total occupancy probability for the $k$th species (across all four orientations) to be
\begin{equation}
	C^{(k)}_n(i, j) =  R^{(k)}_n(i, j) + L^{(k)}_n(i, j) + U^{(k)}_n(i, j) + D^{(k)}_n(i, j), 
\end{equation}
and the total occupancy probability across \textit{all} species to be 
\begin{equation}
	C^\text{tot}_n(i, j) = \sum_{k = 1}^M C^{(k)}_n(i, j).
	\label{eqn:ctot_definition}
\end{equation}

We perform the analysis for the multi-species case parallel to the previous analyses, except we now consider $\vec{P}^{(k)}_n(i, j), P^{(k)}, $ and $\varphi{(k)}$ for each individual species $k$. In the continuum limit, we obtain a system of coupled advection-diffusion PDEs for $1 \leq k \leq M$ described generally by 
\begin{align}
	\dfrac{\partial u^{(k)}}{\partial t} &= \vec{\nabla}\cdot\left[\mathcal{D}^{(k)}(\vec{u})\vec{\nabla}u^{(k)}\right] - \vec{\nabla} \cdot\left(u^{(k)}\vec{v}^{(k)}(\vec{u})\right)
	\label{eqn:pde_system_general}
\end{align}
\noindent where we have written $\vec{u} = \left(u^{(1)}, \dots, u^{(M)}\right)$ and the diffusivity and velocity field are given respectively by 
\begin{align}
	\mathcal{D}^{(k)}(\vec{u}) &= D^{(k)}\left(\dfrac{2 + \varphi^{(k)}}{2-\varphi^{(k)}}\right)(1 - C^\text{tot}), \label{eqn:diffusivity_multispecies}\\ 
	\vec{v}^{(k)}(\vec{u}) &= -D^{(k)}\left(\dfrac{2 + \varphi^{(k)}}{2 - \varphi^{(k)}}\right)(1 - \varphi^{(k)}) \vec{\nabla}C^\text{tot}, \label{eqn:velocity_multispecies}
\end{align}
with
\begin{align}
	D^{(k)} &= \lim_{\Delta, \tau \to 0} \dfrac{P^{(k)}\Delta^2}{4\tau},
\end{align}
where for brevity we have written $C^\mathrm{tot}$ as defined in (\ref{eqn:ctot_definition}).

The diffusivity term $\mathcal{D}^{(k)}$ is similar to that obtained for a single species, except that it is now always a decreasing linear function in the total density $C^\text{tot}$ regardless of the value of $\varphi$. In contrast to the case of a single species, each individual species is now subject to a velocity field $\vec{v}^{(k)}$ oriented parallel to $-\vec{\nabla} C^\text{tot}$, i.e. along the direction of steepest decrease in total agent density. This is consistent with the exclusion effect exerted by the total population on agents of each individual species. As expected, by setting $M = 1$ we recover the single-species case (\ref{eqn:pde_1param_full_eqn}, \ref{eqn:diffusivity_single_species}), and we note that in the non-persistent case $\varphi = 0$, we recover a previously known result derived by Simpson et al. \cite{Simpson2009b}.

\subsection{Incorporating global drift}

For the two-dimensional model with global drift described in Section \ref{sec:globaldrift}, the analysis of Sections \ref{sec:singlespecies} and \ref{sec:multispecies} require some key modifications due to the drift. A discrete master equation is formulated which, under the typical analysis, we convert into continuous variables $(x, y, t)$ and take Taylor expansions. As previously, a transformation (\ref{eqn:transformation_matrix}) into the natural coordinates of the system is then performed. With the presence of drift, we can no longer apply the scaling arguments presented in (\ref{eqn:scaled_phi_def}) since the presence of drift pushes the system away from isotropy. Instead, we put forward an ansatz for the form of $\mathbf{\Phi}$ for the current system to be
\begin{equation}
	\vec{\Phi} = \begin{bmatrix} \phi_1 \\ L_2 + \Delta \phi_2 \\ L_3 + \Delta \phi_3 \\ L_4 + \Delta \phi_4 \end{bmatrix},
	\label{eqn:big_ansatz}
\end{equation}
where $L_2, L_3, L_4$ are non-negative functions of $\phi_1$ only, and $\phi_k = \phi_k(x, y, t), 1 \leq k \leq 4$ are functions in $(x, y, t)$. We next consider the case of a single isolated agent, in which all attempted moves are successful, and formulate the transitions between orientations as a discrete-time Markov chain, for which the state space comprises of the four available orientations $\{R, L, U, D\}$. This is described by the transition matrix
\begin{align}
    \vec{A} = &\dfrac{\lambda}{4} \begin{bmatrix}
       1 + \varphi & 1 - \varphi & 1 & 1 \\
       1 - \varphi & 1 + \varphi & 1 & 1 \\
       1 & 1 & 1 + \varphi & 1 + \varphi \\
       1 & 1 & 1 - \varphi & 1 - \varphi 
    \end{bmatrix}  \nonumber \\ 
   &\quad + \dfrac{1 - \lambda}{4} \begin{bmatrix}
        1 + h & 1 + h & 1 + h & 1 + h\\
        1 - h & 1 - h & 1 - h & 1 - h\\
        1 + v & 1 + v & 1 + v & 1 + v\\
        1 - v & 1 - v & 1 - v & 1 - v
    \end{bmatrix}.
\end{align}
Solving for the stationary distribution $\vec{\pi}$ of the system described by $\vec{A}$ yields the following, where entries correspond respectively to the proportions of time spent by the walker in right, left, up and down orientations under an ergodic interpretation:
\begin{equation}
    \vec{\pi} = \begin{bmatrix}
        \dfrac{-2 + 2h(-1  +\lambda) + \lambda\varphi}{-8 + 4 \lambda \varphi}\\
        \dfrac{1}{4} + \dfrac{h(1-\lambda)}{-4 + 2 \lambda \varphi}\\
        \dfrac{-2 + 2v(-1 + \lambda) + \lambda \varphi}{-8 + 4 \lambda \varphi}\\
        \dfrac{1}{4} + \dfrac{v(1 - \lambda)}{-4 + 2 \lambda \varphi}
    \end{bmatrix}.
\end{equation}
We now suppose that the stationary distribution $\vec{\pi}$ applies also to the case of interacting agents. Noting that $\phi_1$ corresponds to the total occupancy probability across all orientations, we scale $\vec{\pi}$ by $\phi_1$ to obtain in absolute terms the limiting occupancy probabilities for each orientation, $\vec{\pi}^* = \phi_1 \vec{\pi}$, and so we have that
\begin{align}
	\begin{split}
		L_2 &= (\pi^*_1 + \pi^*_2) - (\pi^*_3 + \pi^*_4) = 0,\\
		L_3 &= \pi^*_1 - \pi^*_2 = \dfrac{h ( 1 - \lambda )}{2 - \lambda \varphi}\phi_1, \\
		L_4 &= \pi^*_3 - \pi^*_4 = \dfrac{v( 1 - \lambda)}{2 - \lambda \varphi}\phi_1. 
	\end{split}
\end{align}
We thus arrive at an ansatz for the form of $\vec{\Phi}$,
\begin{equation}
    \vec{\Phi} = \begin{bmatrix}
        \phi_1 \\
        \Delta \phi_2 \\
        \dfrac{h ( 1 - \lambda )}{2 - \lambda \varphi}\phi_1 + \Delta \phi_3 \\
        \dfrac{v( 1 - \lambda)}{2 - \lambda \varphi}\phi_1 + \Delta \phi_4
    \end{bmatrix}.
    \label{eqn:full_ansatz}
\end{equation}

In order to take the continuum limit, we must further enforce a scaling for weak drift \cite{Simpson2009, Simpson2009b} i.e. $h, v \in \mathcal{O}(\Delta)$ to ensure that drift-related terms remain well-behaved in the limit,
\begin{equation}
	h = \Delta H, \qquad v = \Delta V.
\end{equation}
Carrying out the remaining analysis using the ansatz (\ref{eqn:full_ansatz}) yields the resulting advection-diffusion PDE in the continuum-limit for the single species case,
\begin{align}
	\dfrac{\partial u}{\partial t} = \vec{\nabla}\cdot\left[\mathcal{D}(u) \vec{\nabla}u\right] - \vec{\nabla}\cdot\left(u \vec{v}(u)\right)
	\label{eqn:advection_diffusion_single_species}
\end{align}
with diffusivity and velocity field given by  
\begin{align}
	\mathcal{D}(u) &= D\left( \dfrac{2 + \lambda \varphi}{2 - \lambda\varphi} \right)(1 - \lambda\varphi u), \label{eqn:diffusivity_single_species_drift} \\
	\vec{v}(u) &= \dfrac{4D(1 - \lambda)}{2 - \lambda \varphi}(1 - u) \begin{bmatrix} H \\ V \end{bmatrix}, \label{eqn:velocity_single_species_drift}
\end{align}
where $D$ is as defined in Section \ref{sec:singlespecies}. We note that upon setting $\lambda = 1$, we recover the previously presented result (\ref{eqn:diffusivity_single_species}), corresponding to zero drift. For nonzero $\lambda$, we obtain also a velocity field $\vec{v}$ representing the drift velocity parallel to $(H, V)$. Further, the magnitude of each component of $\vec{v}$ is an increasing function of $\varphi$, indicating that motion-persistence reinforces the presence drift.

In the multispecies case, applying the arguments in Section \ref{sec:multispecies} yields a system of advection-diffusion PDEs described by (\ref{eqn:pde_system_general}) with diffusivity and velocity field respectively given by
\begin{align}
	\mathcal{D}^{(k)}(\vec{u}) &= D^{(k)}\left( \dfrac{2 + \lambda^{(k)}\varphi^{(k)}}{2 - \lambda^{(k)}\varphi^{(k)}}\right)\left(1 - C^\text{tot}\right), \\
	\begin{split}
	\vec{v}^{(k)}(\vec{u}) &= D^{(k)}\left\{ \dfrac{4(1 - \lambda^{(k)})}{2 - \lambda^{(k)}\varphi^{(k)}} (1-C^\text{tot}) \begin{bmatrix} H \\ V \end{bmatrix} \right. \\
	&\quad\left.  - \left( \dfrac{2 + \lambda^{(k)}\varphi^{(k)}}{2 - \lambda^{(k)}\varphi^{(k)}} \right) \left(1 - \lambda^{(k)}\varphi^{(k)}\right) \vec{\nabla} C^\text{tot} \right\},\end{split}
\end{align}
where $1 \leq k \leq M$ for $M$ interacting species and $D^{(k)}$ is as defined previously in Section \ref{sec:multispecies}). For brevity, we have written $C^\mathrm{tot}$ as defined in (\ref{eqn:ctot_definition}). We note that the velocity field $\vec{v}^{(k)}$ now contains a term in the direction of $-\vec{\nabla}C^\text{tot}$ arising from population exclusion as well as the drift term parallel to $(H, V)$. Setting $\lambda = 1$ to eliminate drift, the diffusivity and velocity field revert to the previously obtained results (\ref{eqn:diffusivity_multispecies}, \ref{eqn:velocity_multispecies}) as expected.

\section{Results}\label{sec:results}

\begin{figure}
    \centering
    \includegraphics[width = \linewidth]{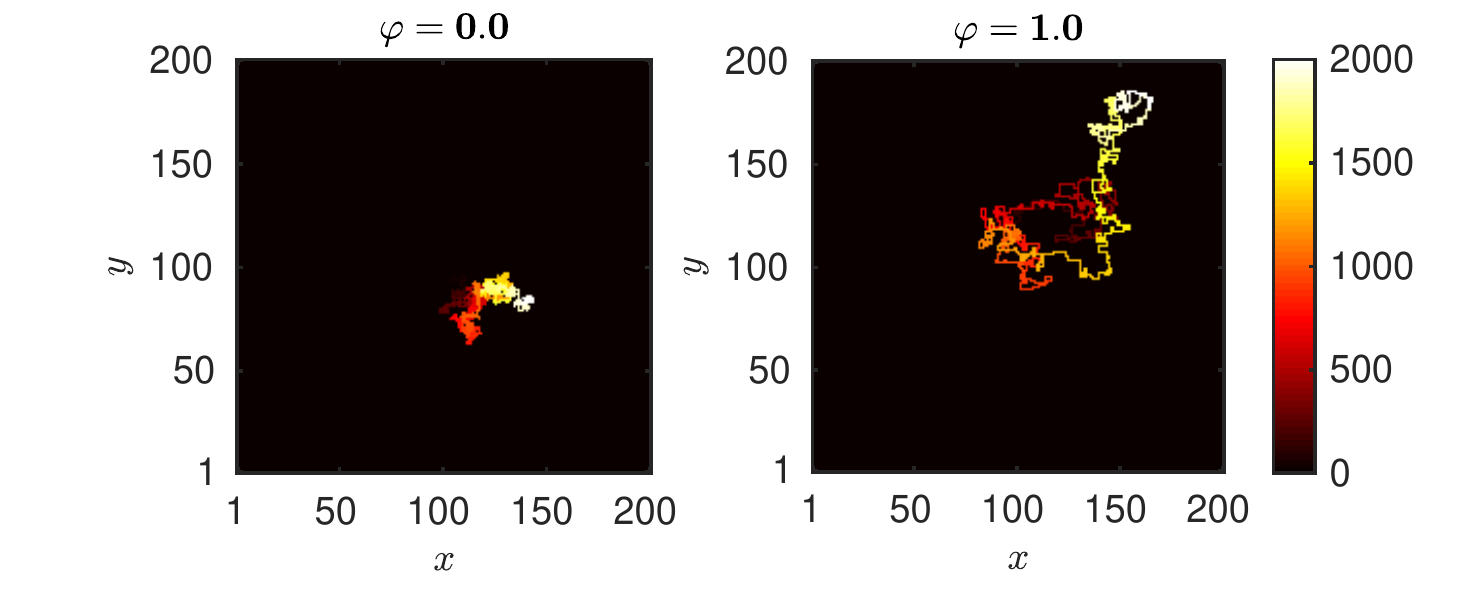}
    \caption{Representative trajectories of a single isolated walker for $0\leq t \leq 2000$ in absence of persistence ($\varphi = 0$) and with maximum persistence ($\varphi = 1$). Color scale shows time evolution.}
    \label{fig:walker_trajectory}
\end{figure}

The agent-based model was implemented for both the one and two-dimensional cases. To obtain qualitative insight on the effect of introducing motion persistence for the case of a single isolated walker, the agent-based simulation was run up to $t = 2000$ on a $200\times 200$ lattice, starting at position $(100, 100)$ with parameters $P = 1, \varphi = 0$ (non-persistent), $\varphi = 1$ (persistent). Results displayed in Fig.~\ref{fig:walker_trajectory} show that the non-persistent walker remains relatively close to its starting point, moving in both a locally and globally random fashion. In contrast, the persistent walker covers much more ground, moving in locally directed bursts that are interspersed with changes in direction, reflecting the local persistence in the presence of the globally unbiased motion characteristic of the persistent random walk \cite{Gavagnin2018, Treloar2011, Codling2008}. 

\subsection{Single species}\label{sec:resultsA}

\begin{figure}
    \centering
    \subfloat[]{\includegraphics[width = \linewidth]{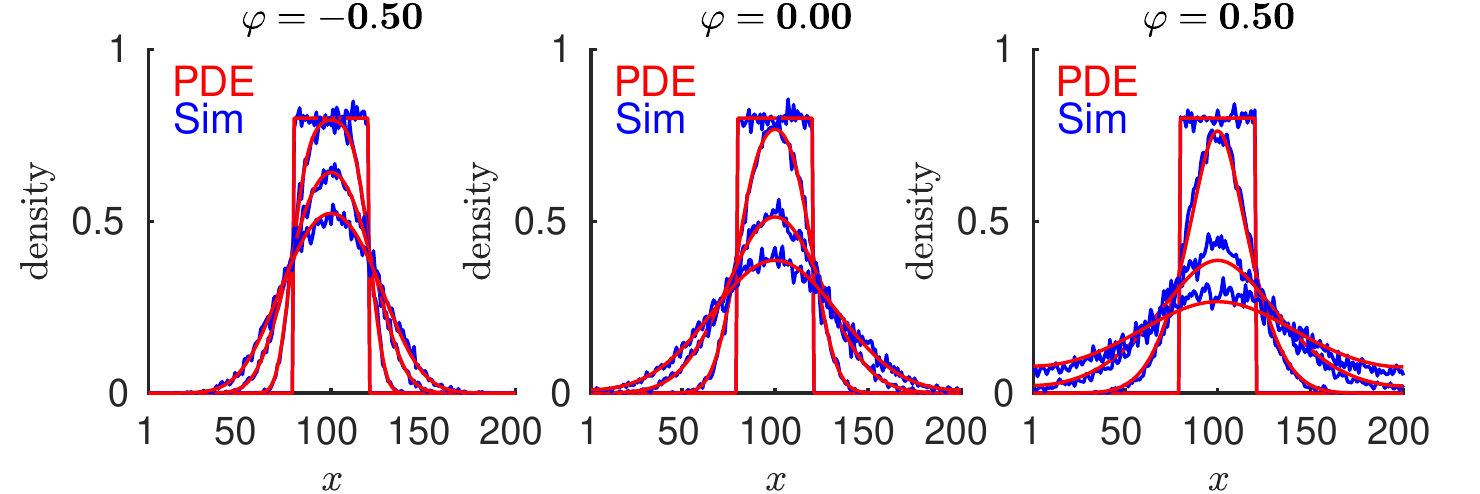}}\\
    \subfloat[]{
	    \includegraphics[width = 0.333\linewidth]{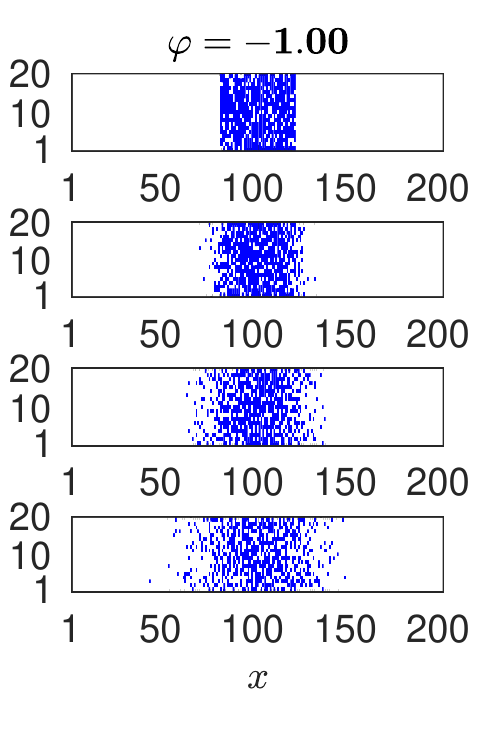}
            \includegraphics[width = 0.333\linewidth]{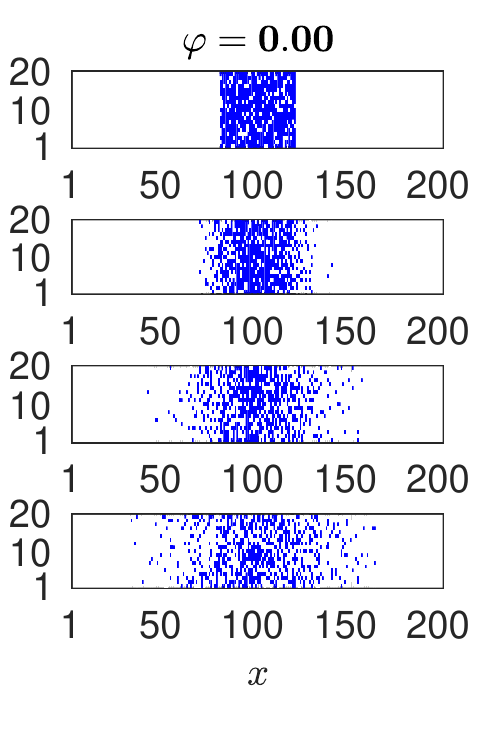}
            \includegraphics[width = 0.333\linewidth]{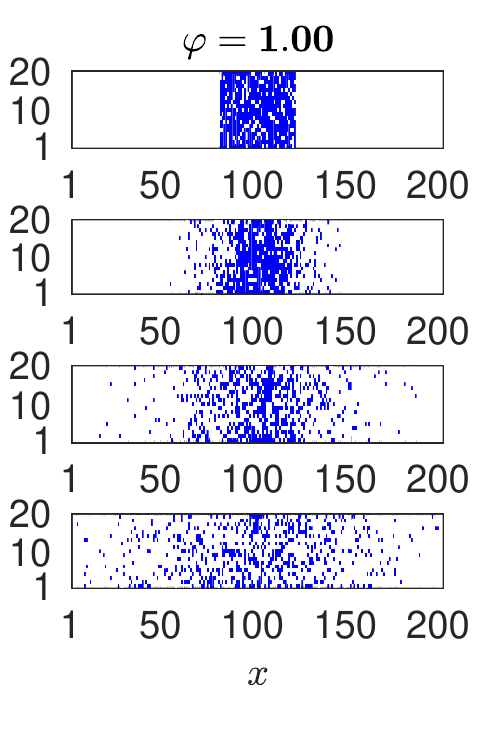}
    }\\
    \subfloat[]{\includegraphics[width = \linewidth]{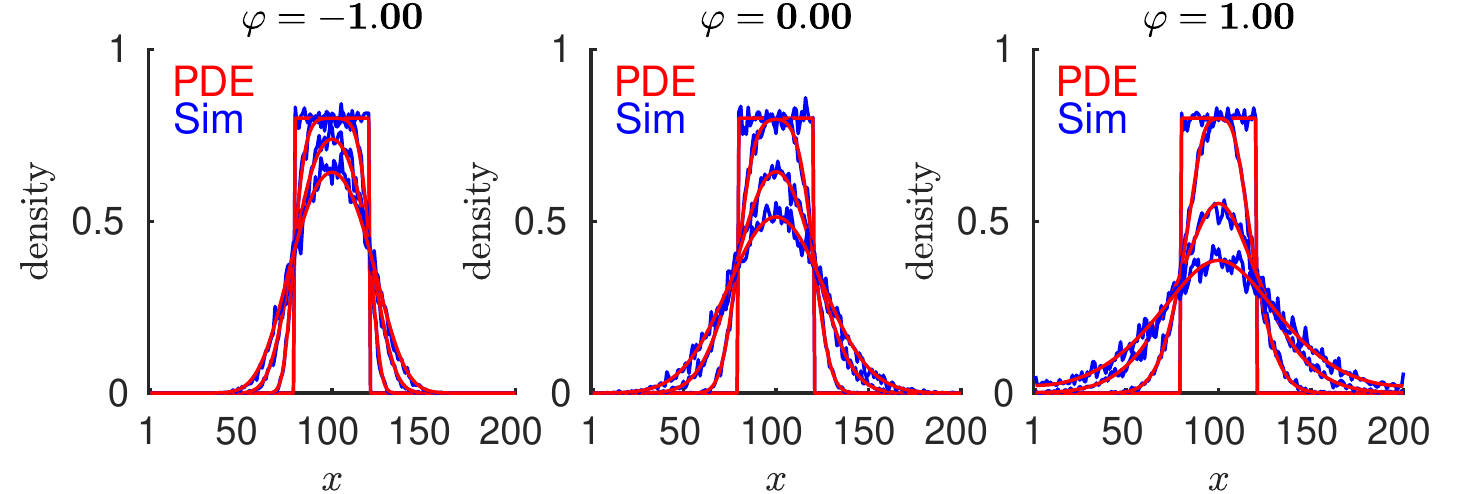}}\\
	\caption{(a) Comparison of simulation and PDE results for one-dimensional model, for non-persistent ($\varphi = 0$) and persistent ($\varphi = -0.5, 0.5$) walkers. (b) Representative snapshots of simulations for two-dimensional model for non-persistent ($\varphi = 0$) and persistent ($\varphi  = -1, 1$) walkers. (c) Comparison of simulation and PDE results for two-dimensional model using column-averaged data for each value of $\varphi$. Results (a--c) shown at times $t = 0, 100, 500, 1000$.}
    \label{fig:singlespecies_plots}
\end{figure}

\begin{figure}
    \centering
    \subfloat[]{\includegraphics[width = 0.5\linewidth]{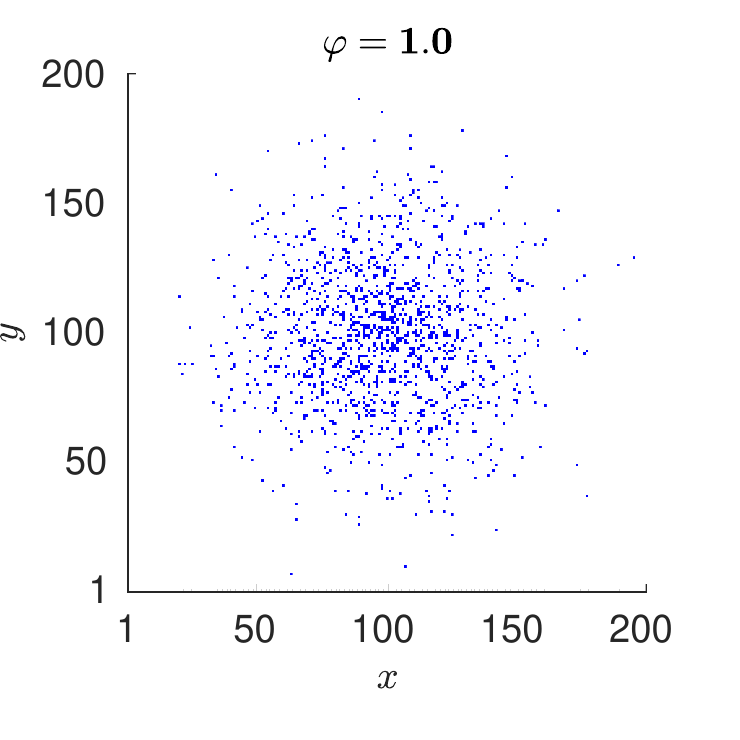}}
    \subfloat[]{\includegraphics[width = 0.5\linewidth]{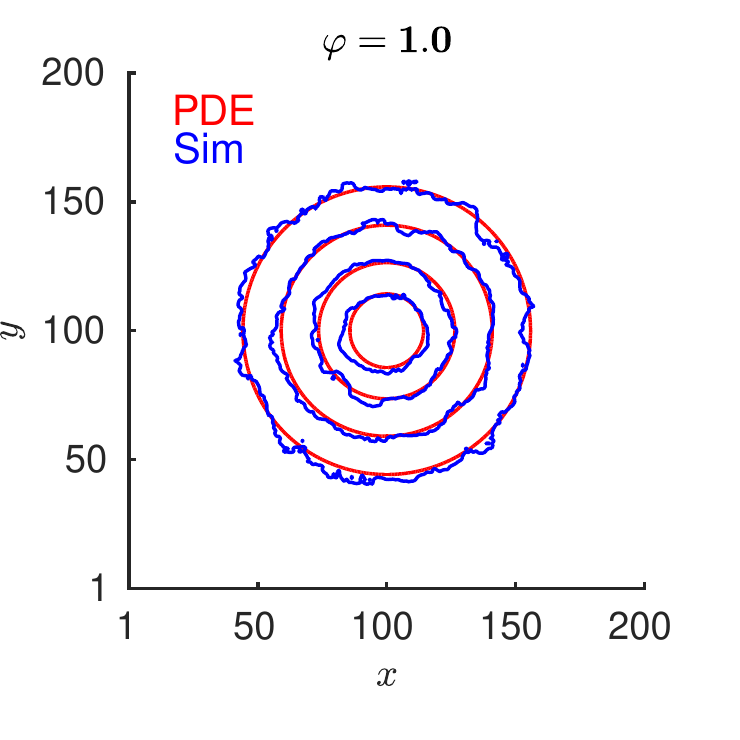}}\\
	\caption{(a) Snapshot of a single representative simulation at $t = 1000$ for the two-dimensional model on a square lattice with $\varphi = 1$. (b) Contour plot of simulation results against PDE solutions at $t = 1000$, with levels shown at 10\%, 25\%, 50\% and 75\% of maximum simulated occupancy.}
    \label{fig:single_species_2d}
\end{figure}
In this subsection, we examine the results of our continuum-limit analysis for homogeneous populations of motion-persistent walkers. All simulations were performed up to $t = 1000$ and sampled at $t = 0, 100, 500, 1000$, for which corresponding PDE solutions were found. For the one-dimensional case, we conducted simulations on a lattice of size $200$, starting with an initially uniform agent distribution at density 0.8 on $[80, 120]$ for both non-persistent ($\varphi = 0$) and persistent ($\varphi = \pm0.5$) walkers, and $P = 1$. For all simulations in this work, initial agent orientations were randomly assigned with uniform probability from two and four possible orientations for one and two dimensions respectively, in keeping with the symmetry assumption of Section \ref{sec:singlespecies}. Simulations were averaged over 500 realisations, and the continuum-limit PDE (\ref{eqn:pde_1param_full_eqn}, \ref{eqn:pde_1d_1param_full_diffusivity}) was solved using the \texttt{pdepe} routine offered by \textsc{Matlab} with default settings, subject to the corresponding initial conditions and no-flux boundary conditions. The results shown in Fig.~\ref{fig:singlespecies_plots}(a) confirm a close agreement between the agent-based discrete model and its continuum-limit approximation. 
From these results, we note that a positive persistence parameter $\varphi = 0.5$ results in faster dispersal of agents from the initial configuration, compared with the non-persistent case where $\varphi = 0$. This is concordant with our observations made earlier on the motion of isolated agents. Conversely, $\varphi = -0.5$ enforces a negative persistence and we observe that agents disperse more slowly as a result --- this agrees with the fact that agents are now averse to moving in the same direction for multiple steps. As expected from Section \ref{sec:singlespecies}, the match between simulation and continuum results deteriorated for large $\varphi$ and high densities (not shown) and eventually solution of the PDE (\ref{eqn:pde_1param_full_eqn}, \ref{eqn:pde_1d_1param_full_diffusivity}) failed for sufficiently large $\varphi$ since a negative diffusivity resulted. Solutions for higher values of $\varphi$ can be accommodated by starting with a sufficiently low initial density. 

We now turn our attention to the two-dimensional case. For ease of visualisation, simulations were conducted on a $200\times 20$ lattice with $y$-invariant initial conditions and lattice occupancies were averaged along columns \cite{Gavagnin2018, Simpson2009b}. For the PDE model, computing a column-averaged density amounts to integrating out $y$-dependence. We find that the column-averaged density satisfies the same PDE as the original density when we drop $y$-terms in the gradient operator, i.e. we take $$\vec{\nabla} \equiv \dfrac{\partial}{\partial x}$$ for column-averaged solutions. Agents were initially uniformly distributed on $[80, 120]\times[1, 20]$ at density 0.8. Parameters corresponding to non-persistent ($\varphi = 0$) and persistent ($\varphi = \pm 1$) walkers were used, and $P = 1$ held fixed. Simulations were averaged across 20 realisations, and solutions to the column-averaged PDE (\ref{eqn:pde_1param_full_eqn}, \ref{eqn:diffusivity_single_species}) were found as previously. 
{Snapshots of the lattice agent distribution taken at each time for a single simulation realisation for each value of $\varphi$ are shown in Fig.~\ref{fig:singlespecies_plots}(b), and a comparison of column-averaged simulation results to PDE solutions are presented in Fig.~\ref{fig:singlespecies_plots}(c).} As previously noted, the column-averaged results provide an excellent match to simulation. A positive persistence parameter ($\varphi = 1$) again results in faster dispersal of agents compared to the non-persistent ($\varphi = 0$) case, whilst a negative persistence parameter ($\varphi = -1$) has the opposite effect. In particular, we note that whilst in the one-dimensional case the PDE solution deviates from simulated results for large values of the persistence parameter and eventually becomes badly behaved, in two dimensions neither of these issues are apparent. This suggests that the additional degree of freedom has non-negligible consequences for the behaviour of the system, and that a level of caution may be necessary when attempting to generalise one-dimensional results into higher dimensions. 

Subsequently we employed a $200\times 200$ lattice and an initial agent distribution on $[80, 120]\times [80, 120]$ at density 0.8 to investigate the correspondence between discrete and continuum results in a fully two-dimensional case. Parameters $\varphi = 1, P = 1$ were used. Results at $t = 1000$ were averaged across 20 simulation realisations and smoothed using a $10\times 10$ convolution kernel $\vec{J}_{10}$. Solutions to the PDE (\ref{eqn:pde_1param_full_eqn}, \ref{eqn:diffusivity_single_species}) for $t = 1000$ were found numerically using the \texttt{solvepde} routine from the \textsc{Matlab} PDE Toolbox. Fig.~\ref{fig:single_species_2d} shows contour plots of both the simulated and continuum density at 10\%, 25\%, 50\% and 75\% of the maximum simulated occupancy and again a close match confirms that the derived continuum description accurately captures population-level behaviour in a general two dimensional scenario.

\begin{figure}
    \centering
        \subfloat[]{
		\includegraphics[width = 0.33\linewidth]{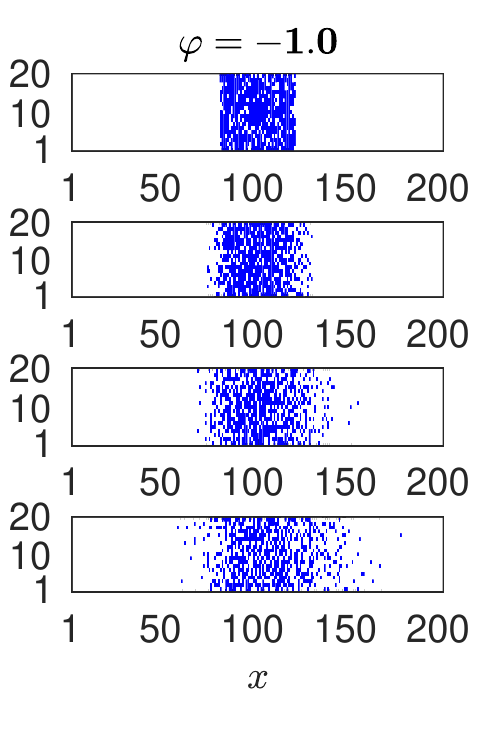}
		\includegraphics[width = 0.33\linewidth]{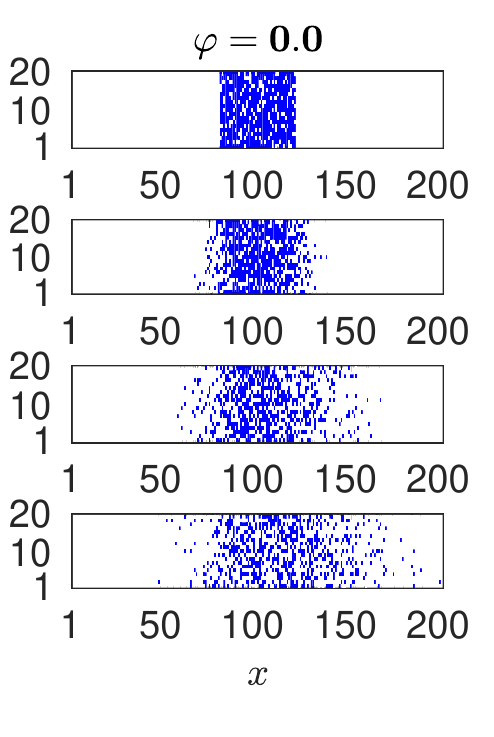}
		\includegraphics[width = 0.33\linewidth]{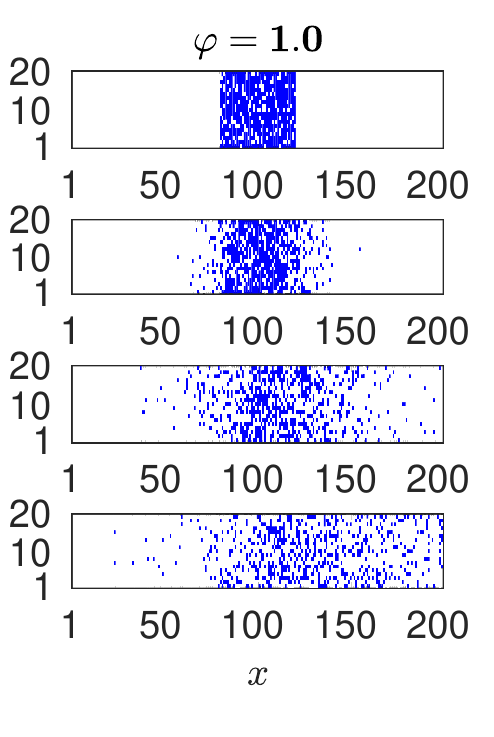}
	}\\
        \subfloat[]{\includegraphics[width = \linewidth]{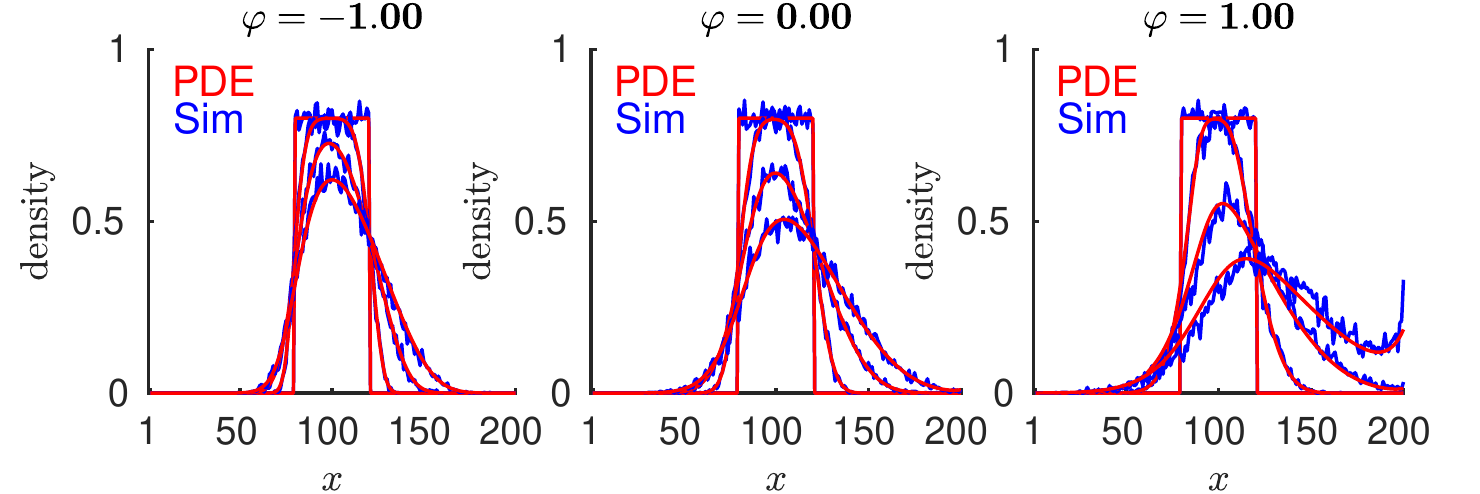}}\\
	\caption{(a) Representative snapshots of simulations for non-persistent ($\varphi = 0$) and persistent ($\varphi = -1, 1$) walkers, and $\lambda = 0.9, h = 0.5, v = 0$. (b) Comparison of simulation and PDE results for two-dimensional model with global drift using column-averaged data for each value of $\varphi$. Results (a, b) shown at times $t = 0, 100, 500, 1000$.}
    \label{fig:singlespecies_drift_plots}
\end{figure}

For the two-dimensional model with global drift described in Section \ref{sec:globaldrift}, we performed simulations using a setup identical that used previously for Fig.~\ref{fig:singlespecies_plots}(b, c), except parameters $P = 1, \varphi = -1, 0, 1$ were used and $\lambda = 0.9, h = 0.5, v = 0$ in order to enforce a weak rightwards drift. Column-averaged simulation results were averaged over 20 realisations, and corresponding numerical solutions to the PDE (\ref{eqn:advection_diffusion_single_species}, \ref{eqn:diffusivity_single_species_drift}, \ref{eqn:velocity_single_species_drift}) were found at each time. 
Representative simulation snapshots for each value of $\varphi$ are shown in Fig.~\ref{fig:singlespecies_drift_plots}(a), and a comparison between simulation and continuum results is shown in Fig.~\ref{fig:singlespecies_drift_plots}(b).
From this, the addition of a global drift effect is clearly visible. At long times, agents accumulate near the right boundary due to the imposed no-flux condition. We note that the presence of persistence in the $\varphi = 1$ case has a synergistic effect with the rightwards drift. Thus, from both our agent-based and continuum results, we conclude that presence of a positive persistence effect enhances both the rate of agent dispersal as well as the extent of agent migration in the presence of an external drift. 

\subsection{Multiple interacting species}\label{sec:resultsB}

\begin{figure*}
    \centering
        \subfloat{\includegraphics[width = 0.48\linewidth]{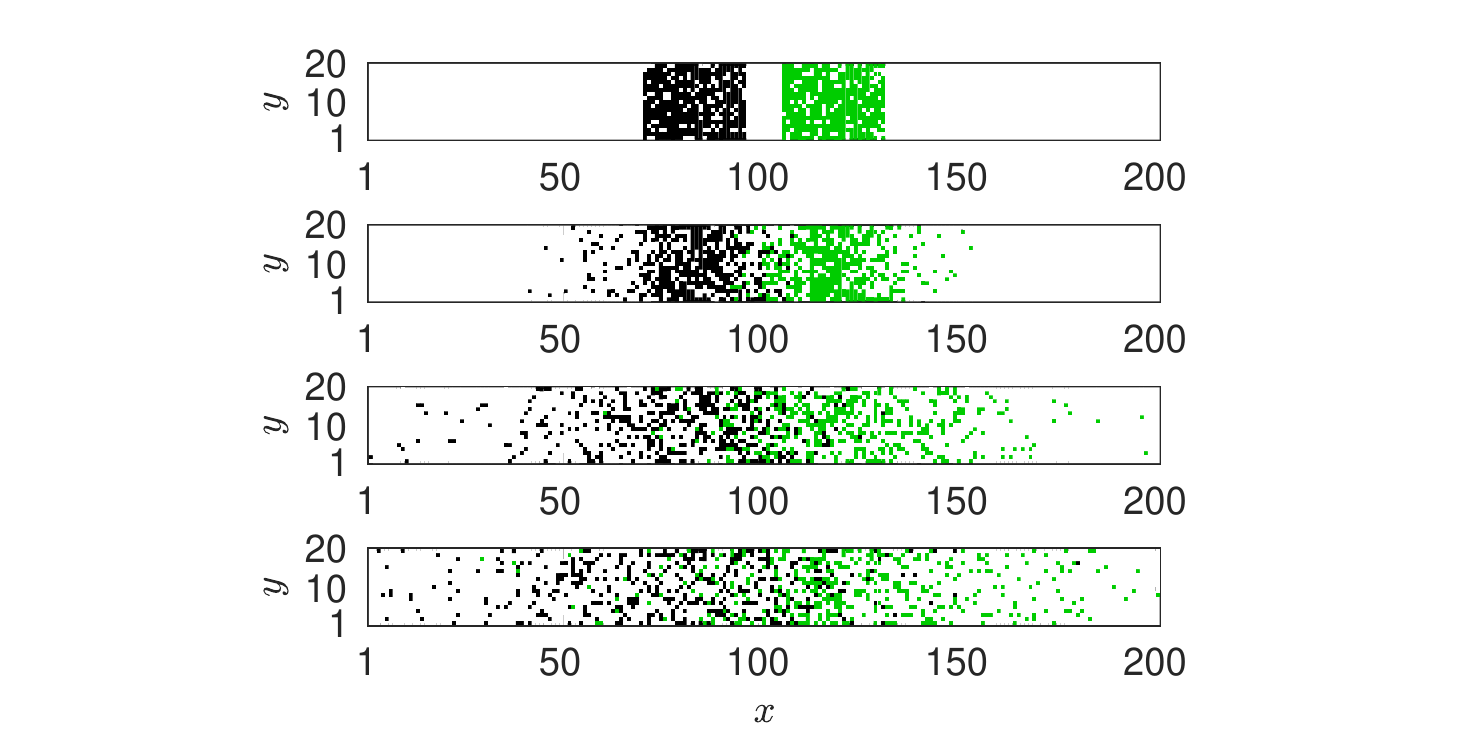}}\addtocounter{subfigure}{-1}
        \subfloat{\includegraphics[width = 0.48\linewidth]{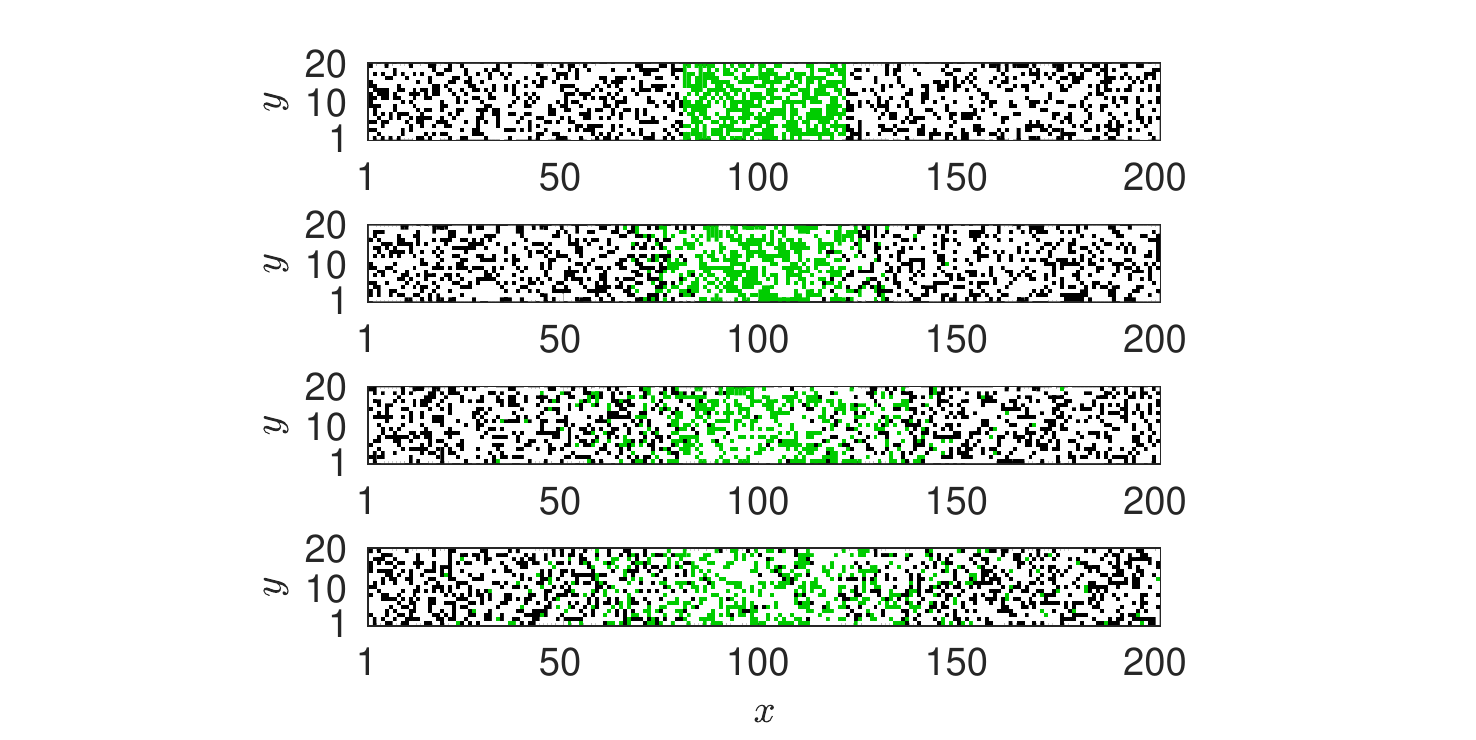}}\addtocounter{subfigure}{-1}\\
        \subfloat[]{\includegraphics[width = 0.48\linewidth]{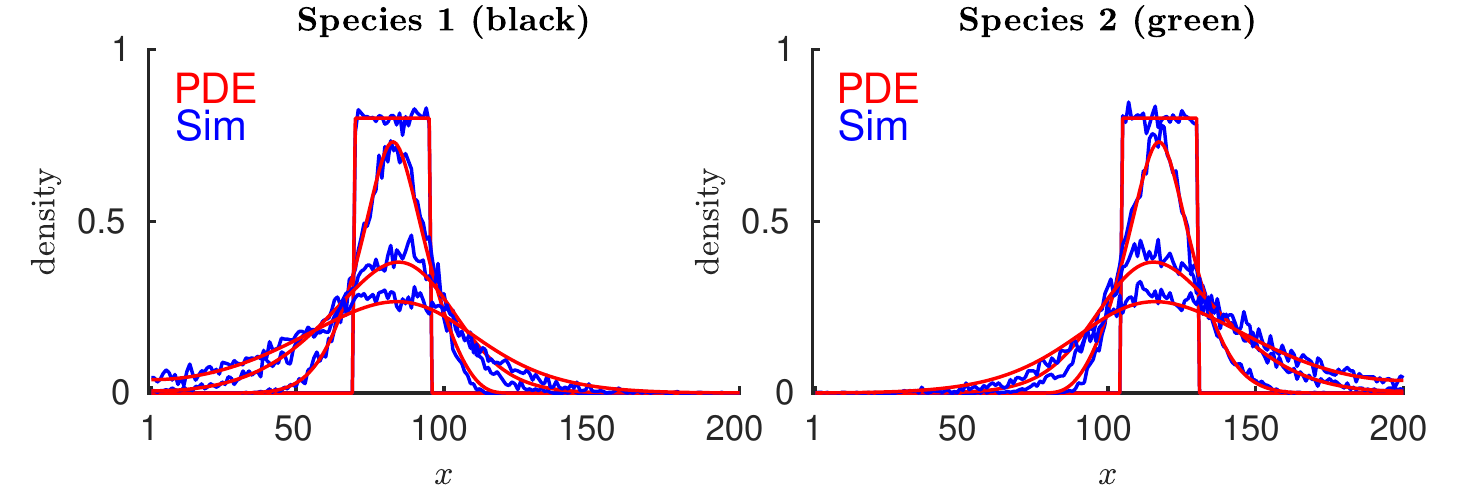}}
        \subfloat[]{\includegraphics[width = 0.48\linewidth]{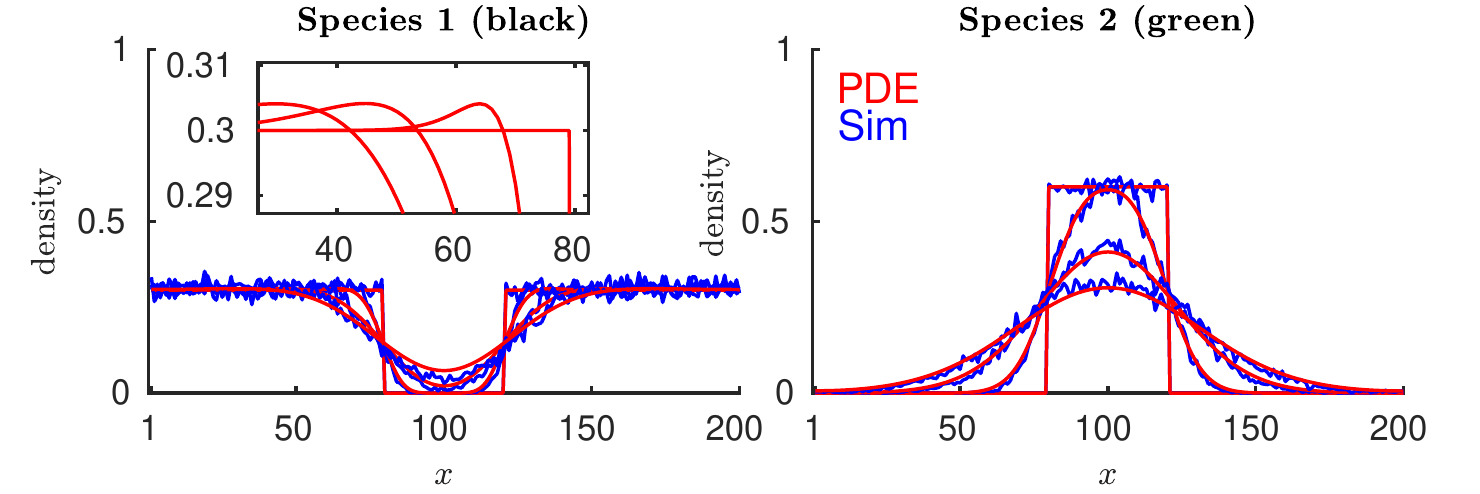}}
		\caption{Comparison of simulation and PDE results for various multi-species problems: (a) two identical species with $P^{(1)} = P^{(2)} = 1, \varphi^{(1)} = \varphi^{(2)} = 1$ dispersing onto the lattice. (b) Mixing of persistent species 2 (green, $P^{(2)} = 1, \varphi^{(2)} = 1$) into sparse population of non-persistent species 1 (black, $P^{(1)} = 1, \varphi^{(1)} = 0$). Results for (a, b) shown at times $t = 0, 100, 500, 1000$. }
    \label{fig:multispecies_plots}
\end{figure*}

\begin{figure}
    \subfloat[]{\includegraphics[width = \linewidth]{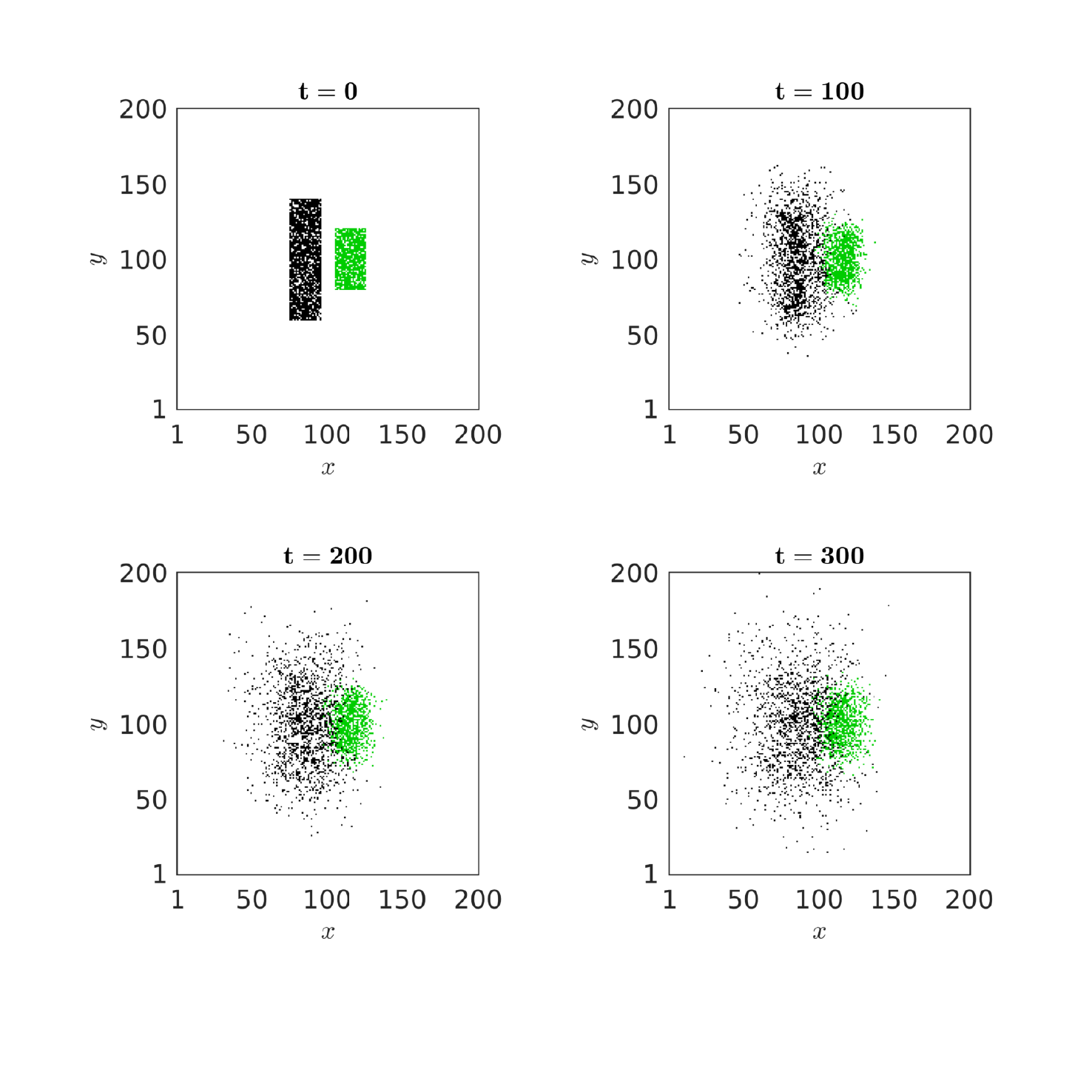}}\\
    \subfloat[]{\includegraphics[width = \linewidth]{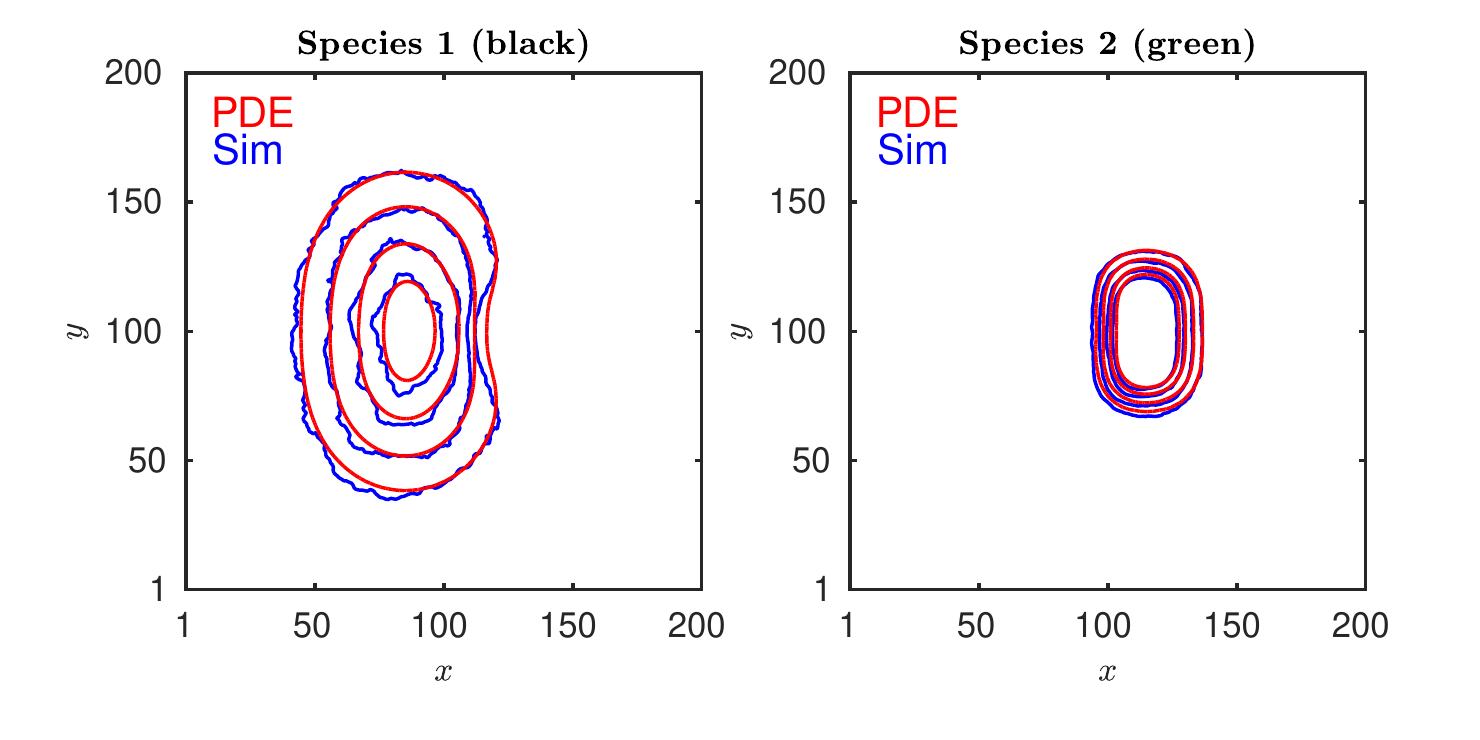}}
	\caption{(a) Representative simulation snapshots shown at $t = 0, 100, 200, 300$ for interaction of fast-moving, persistent species 1 (black, $P^{(1)} = 1, \varphi^{(1)} = 1$) with slow-moving, non-persistent species 2 (green, $P^{(2)} = 0.3, \varphi^{(2)} = 0$) on a square lattice. (b) Comparison of simulation and PDE results. Results shown at $t = 300$ with contours shown at 10\%, 25\%, 50\% and 75\% of the maximum simulated occupancy of species 1.}
     \label{fig:multispecies_plots_2d}
\end{figure}

\begin{figure}
        \subfloat[]{\includegraphics[width = \linewidth]{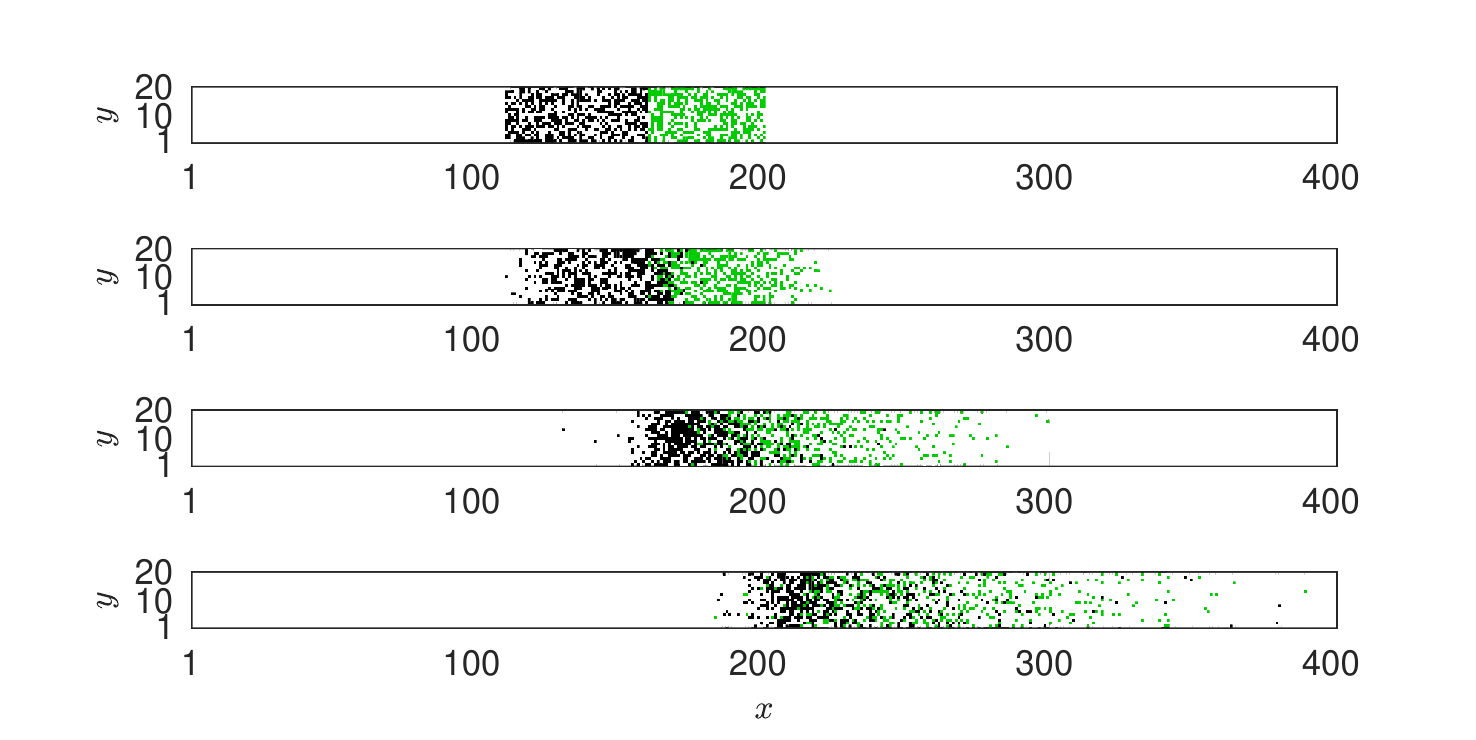}}\\
	\subfloat[]{\includegraphics[width = \linewidth]{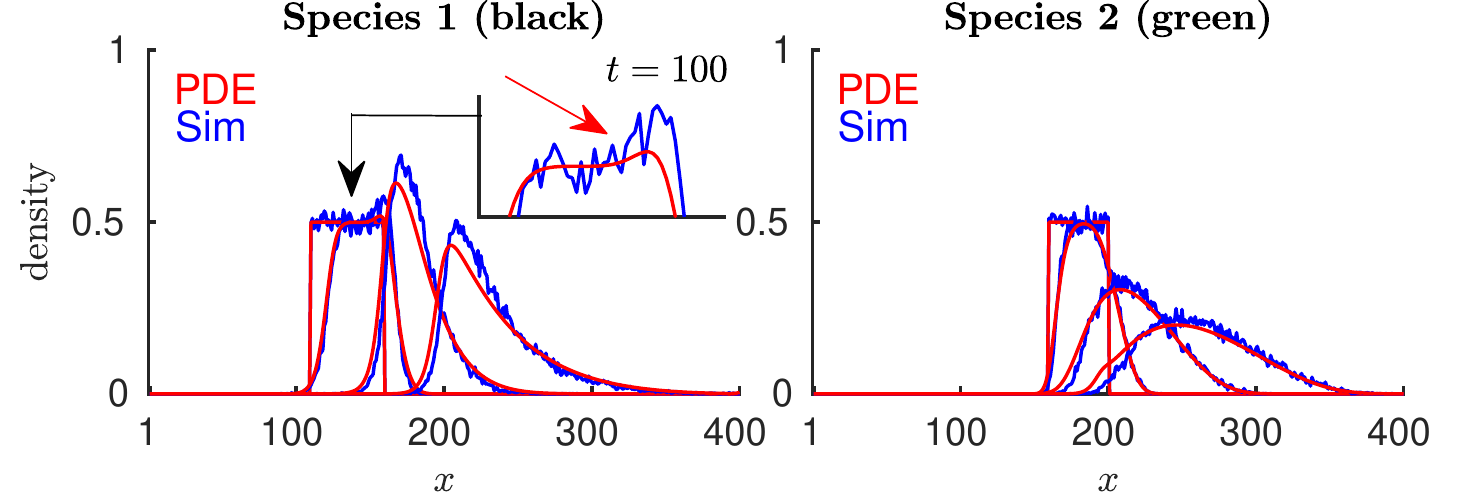}}
	\caption{(a) Representative snapshots of simulations for interaction of a persistent drifting species (black, $\varphi^{(1)} = 1, \lambda^{(1)} = 0.75, h^{(1)} = 1, v^{(1)} = 0$) with a non-persistent drifting species (green, $\varphi^{(2)} = 0, \lambda^{(2)} = 0.75, h^{(2)} = 1, v^{(2)} = 0$) on a $400\times 20$ lattice. (b) Comparison of simulation and PDE results. Inset: red arrow indicates formation of a small density front at $t = 100$ resulting from persistence, drift and spatial exclusion. Results (a, b) shown at $t = 0, 100, 500, 1000$.}
        \label{fig:multispecies_drift}
\end{figure}

For the case of two interacting species, we investigated the correspondence between agent-based simulations and solutions to the continuum PDE (\ref{eqn:pde_system_general}, \ref{eqn:diffusivity_multispecies}, \ref{eqn:velocity_multispecies}) for a range of scenarios. As in Section \ref{sec:resultsA}, unless otherwise mentioned, all simulations were performed up to $t = 1000$ and sampled at $t = 0, 100, 500, 1000$. First, we consider a $y$-invariant problem with two persistent species dispersing across a $200\times 20$ lattice. Initially, agents were uniformly distributed at density 0.8 with species 1 on $[70, 95]\times[1, 20]$ and species 2 on $[105, 130]\times[1, 20]$. Both species were subject to identical parameters $P^{(1, 2)} = 1, \varphi^{(1, 2)} = 1$. Simulations results were averaged across $20$ realisations, and column-averaged continuum PDE solutions were found as previously in Section \ref{sec:resultsA}. In Fig.~\ref{fig:multispecies_plots}(a), we show a snapshot of a single realisation of the discrete simulation and then compare simulation to continuum results. From this, we observe that the continuum approximation agrees well with simulation data for each species. In particular, we note that the species distributions become skewed, reflecting the large-scale effect of spatial exclusion between agents of different species at the interface near $x = 100$. 

Following Simpson et al. \cite{Simpson2009b}, we consider a mixing scenario with a non-persistent `filler' species 1 distributed on $([1, 79] \cup [121, 200])\times[1, 20]$ at density 0.3, and a persistent `invading' species 2 distributed on $[80, 120]\times[1, 20]$ at density 0.6. We take $P^{(1, 2)} = 1$ and $\varphi^{(1)} = 0, \varphi^{(2)} = 1$. Simulations were run as for the previous example but results were averaged over 40 realisations. Column-averaged continuum PDE solutions were found as previously in Section \ref{sec:resultsA}, and results are shown in Fig.~\ref{fig:multispecies_plots}(b). From these results, we again confirm that the continuum model provides an accurate description of the agent-based process. As observed by Simpson et al. \cite{Simpson2009b} for non-persistent agents, we also note that continuum density profiles obtained for the mixing problem display nonmonotone behaviour as a result of interactions, as shown in Fig.~\ref{fig:multispecies_plots}(b, inset).

Extending our investigation more generally to a two-dimensional problem, we consider again the $200\times 200$ lattice as previously, now with a fast-moving, persistent species 1 ($P^{(1)} = 1, \varphi^{(1)} = 1$) interacting with a slow-moving, non-persistent species 2 ($P^{(2)} = 0.3, \varphi^{(2)} = 0$), initially uniformly distributed at density 0.8 on $[75, 95]\times[60, 140]$ and $[105, 125]\times[80, 120]$ respectively. Simulations were performed up to time $t = 300$. Results were averaged over 20 realisations and smoothed by convolution as previously. PDE solutions for $t = 300$ were found as in Section \ref{sec:resultsA}. Fig.~\ref{fig:multispecies_plots_2d}(a) shows snapshots of of a single simulation at $t = 0, 100, 200, 300$, and Fig.~\ref{fig:multispecies_plots_2d}(b) shows contours from the PDE and simulated results at $t = 300$ at levels of 10\%, 25\%, 50\% and 75\% of the maximum simulated occupancy of species 1. We find that the continuum model largely captures the simulated behaviour, again confirming that the continuum approximation holds well for two-dimensional systems with interacting species. 

Finally we consider combined drift and multi-species interactions with persistence. We found that in general the continuum approximation incorporating all three effects is more fragile and often breaks down in situations where species distributions have differing drift velocities and thus `collide' with each other. We deal with a $400\times 20$ lattice with a persistent, drifting species 1 ($\varphi^{(1)} = 1$) and a non-persistent drifting species 2 ($\varphi^{(2)} = 0$) initially distributed on $[110, 159]\times [1, 20]$ and $[160, 200]\times [1, 20]$ respectively with density 0.5. Parameters used were $P^{(1, 2)} = 1, \lambda^{(1, 2)} = 0.75, h^{(1, 2)} = 1, v^{(1, 2)} = 0$. Representative simulation snapshots at each time are shown in Fig.~\ref{fig:multispecies_drift}(a). Simulations were then averaged over 40 realisations and Fig.~\ref{fig:multispecies_drift}(b) shows a comparison of the simulation and continuum results. We note that the species 1 distribution is more sharp and pointed compared to species 2, a result of the interplay of persistence, drift, and spatial exclusion. This can also be seen from the `bunching up' of black agents in the snapshot data. In particular, we note that at $t = 100$ a small density front is formed (see inset, indicated by red arrow), an effect qualitatively captured by the continuum description. This effect occurs because motion persistence amplifies drift as observed previously, and disappears when persistence for species 1 is switched off.

Whilst the continuum description has held mostly accurate for the problems considered, we observe that the continuum PDE results tend to deviate from the true simulated results near the interface between species. This is especially evident in Fig.~\ref{fig:multispecies_plots_2d} and  Fig.~\ref{fig:multispecies_drift}, suggesting that non-negligible occupancy correlations at such interfaces may render mean-field assumptions inaccurate. To correct for such effects in these regions, further analytical or computational work may be necessary \cite{Simpson2011, Markham2013, Teomy2019}. 

\section{Inferring persistence from data}

In the following discussion, we investigate the use of pair correlation functions (PCFs) for inferring the presence or absence of persistence in agent movement from simulation results. PCFs are a common tool from statistical mechanics used to characterise the average interactions in a spatial particle distribution, and have recently been applied to the context of examining cellular motility \cite{Johnston2019, Agnew2014, Gavagnin2018b}. We hypothesise that PCFs could be applied to examine motion persistence in experimentally derived data, appropriately discretised to a lattice \cite{Mort2016, Agnew2014, Deroulers2009}. We first consider the case of a swarm of walkers uniformly distributed on a two-dimensional lattice, and then extend discussion to the non-uniform case of a swarm of walkers diffusing from an initial condition. 

Gavagnin, Owen and Yates \cite{Gavagnin2018b} recently provided relevant formulae for computing a on-lattice pair correlation function based on the $\ell_1$ (Manhattan) norm. Following this, we define the pair correlation function for a given distance metric $d$ as
\begin{equation}
	f_d(m) = \dfrac{n_d(m)}{\bar{n}_d(m)}\label{eqn:pcf}
\end{equation}
where $n_d(m)$ is the number of unique particle pairs observed to be separated by distance $m$, and $\bar{n}_d(m)$ is the number of such pairs expected to be observed if the particle distribution was truly random, referred to as complete spatial randomness (CSR) \cite{Agnew2014, Gavagnin2018b}. We consider a rectangular lattice of dimensions $L_x \times L_y$ and enforce periodic boundaries. The Manhattan distance between $(x, y)$ and $(x', y')$ is thus
\begin{align}
\begin{split}
	d(x, y) &= \min \left\{ |y - y'|, L_y - |y - y'| \right\} \\
	&\quad + \min \left\{ |x - x'|, L_x - |x - x'|\right\} \\
	&\leq \dfrac{L_x + L_y}{2} .
\end{split}\label{eqn:mahattan}
\end{align}
\noindent As in \cite{Gavagnin2018b}, we consider
\[
1 \leq m \leq \min \left\{ \left \lfloor{\frac{L_x}{2}} \right \rfloor, \left \lfloor {\frac{L_y}{2}}\right \rfloor \right\},
\]
and employ the normalisation factor for periodic boundaries,
\begin{equation}
	\bar{n}_d(m) = \dfrac{2mN(N-1)}{L_x L_y - 1},
\end{equation}
\noindent where $N$ is the total number of agents present on the lattice. In the following discussion, we take $f(m) = f_d(m)$ to be the PCF calculated using the Manhattan norm.

\subsection{Pair correlation function for a uniform swarm} 

\begin{figure*}
	\begin{center}
		\subfloat[]{\includegraphics[width = \linewidth]{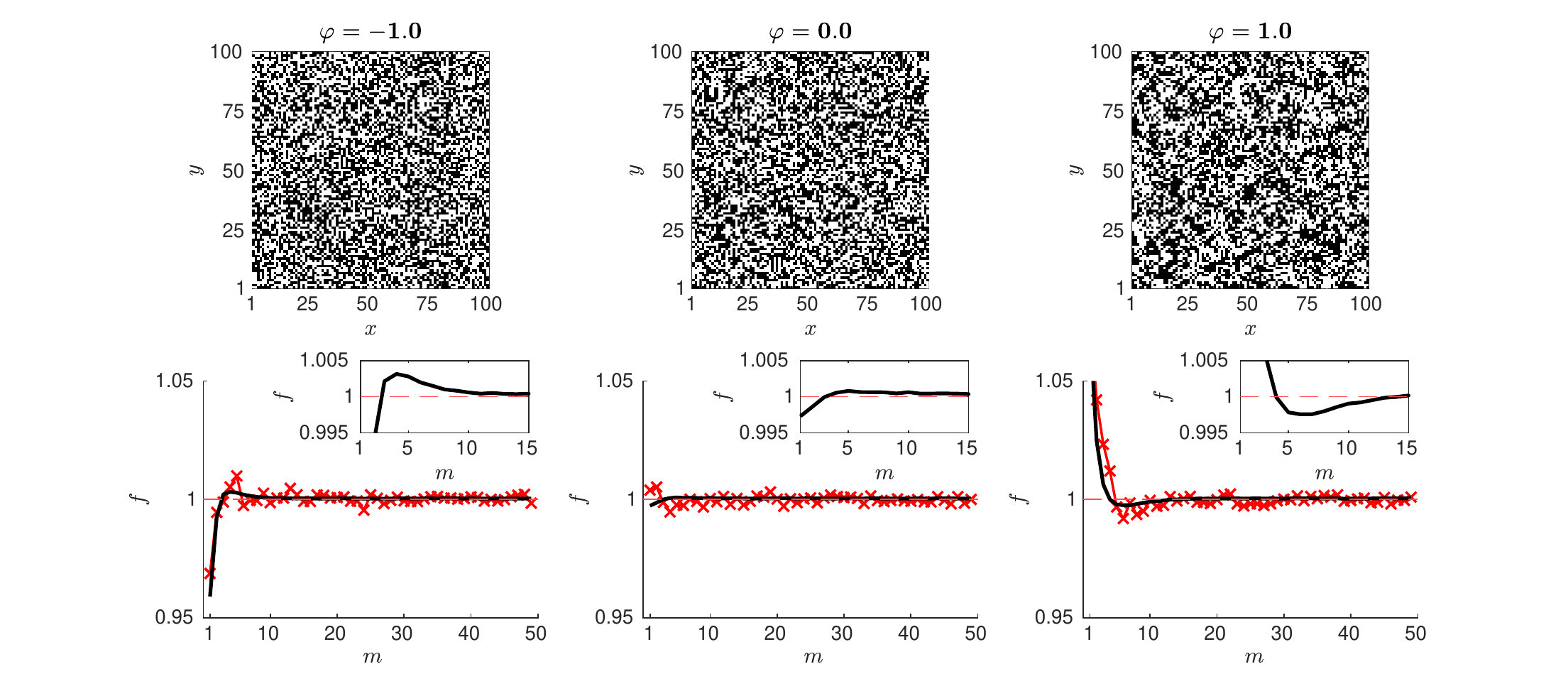}}\\
		\subfloat[]{\includegraphics[width = \linewidth]{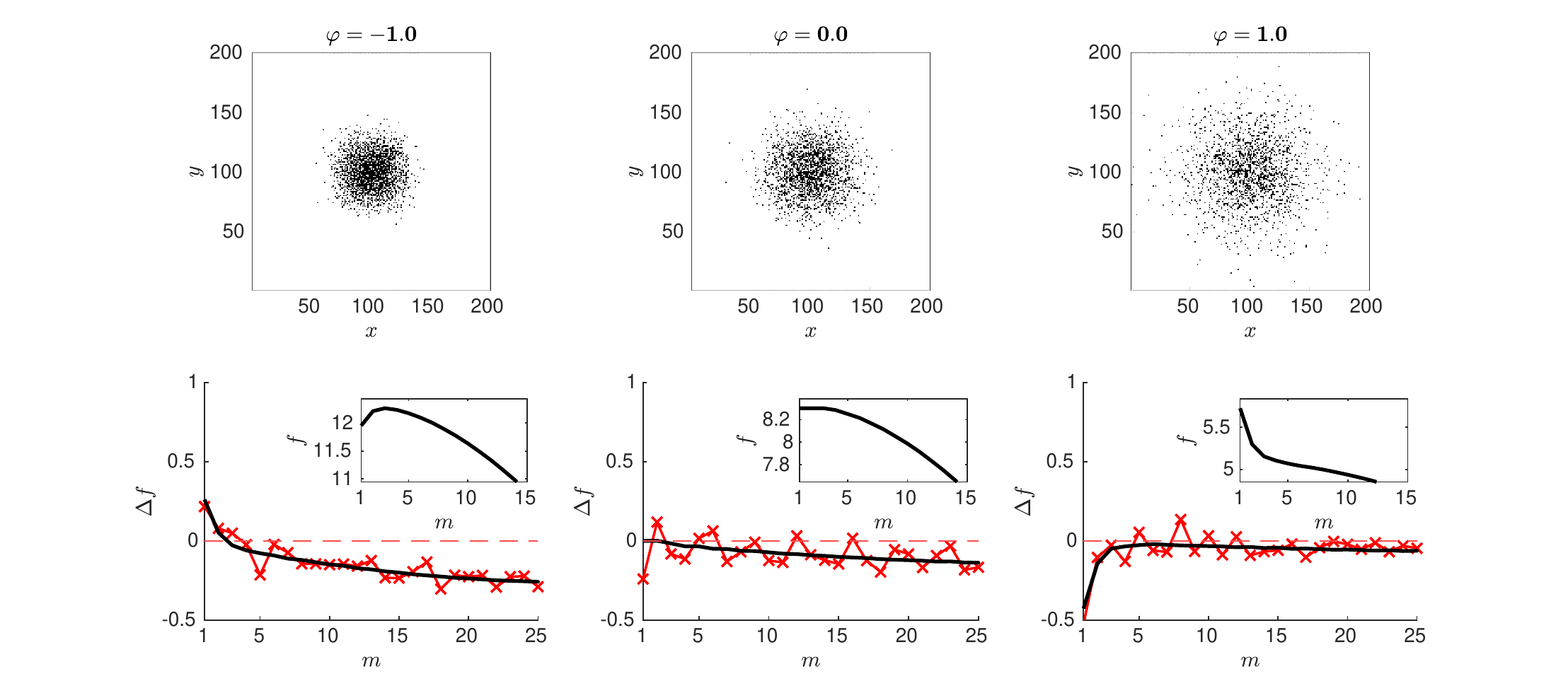}}
	\end{center}
	\caption{(a) Pair correlation function $f(m)$ [see (\ref{eqn:pcf})] calculated for the Manhattan distance [see (\ref{eqn:mahattan})] e computed at $t = 20$ for a $100\times 100$ lattice initially uniformly populated with identical agents at density 0.5, with parameters $P = 1$ and $\varphi = -1, 0, 1$ respectively. (b) $\Delta f = f(m+1) - f(m)$, where $f$ is the pair correlation function as previously, computed at $t = 500$ for diffusing swarms on a $200\times 200$ lattice with parameters $P = 1$ and $\varphi = -1, 0, 1$ respectively. (b, inset) $f$ computed at $t = 500$ for for the same data. PCF results shown both averaged over 1000 simulations (black) and for a single simulation (red). }
	\label{fig:pcf}
\end{figure*}

\begin{figure}
	\begin{center}
		\includegraphics[width = \linewidth]{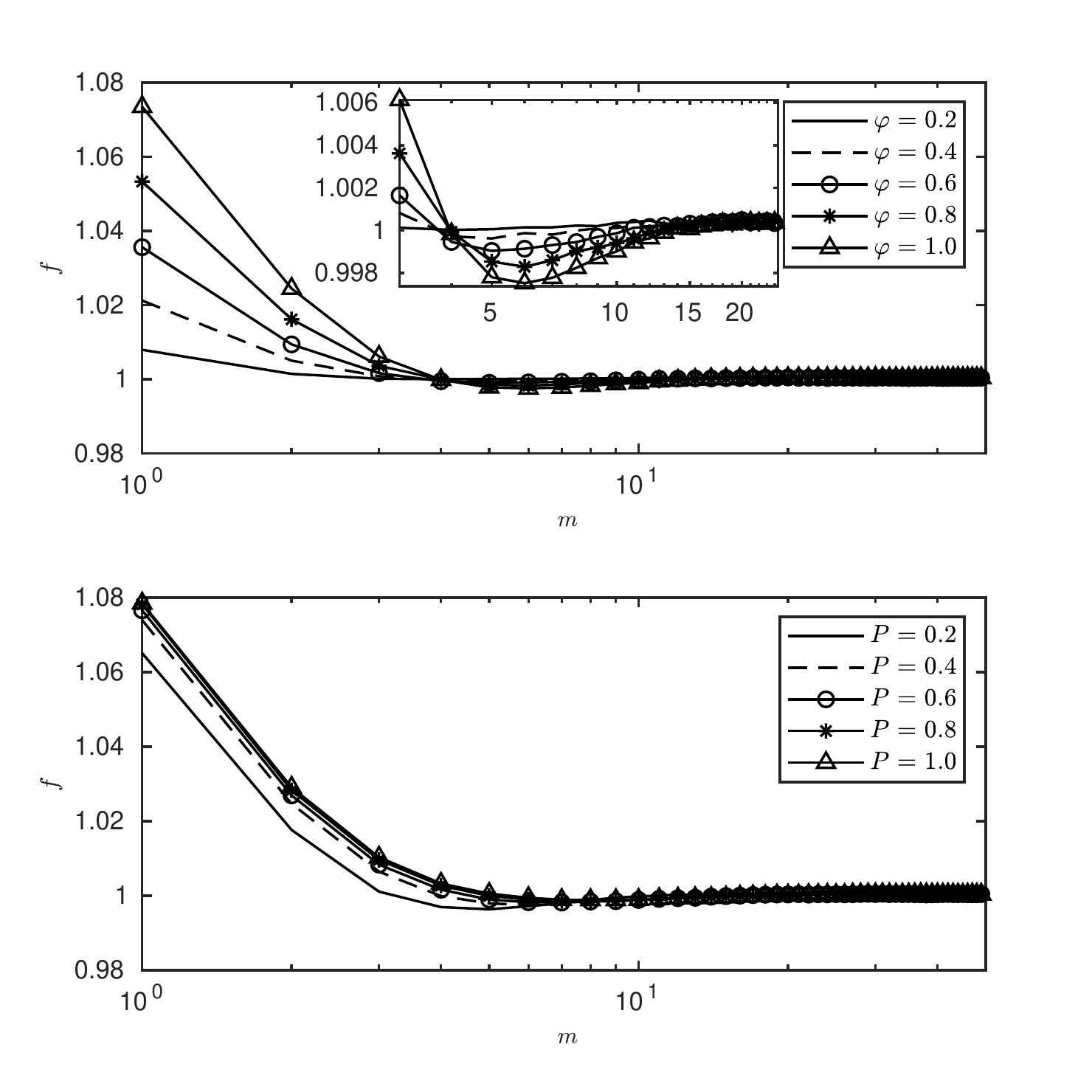}
	\end{center}
	\caption{Comparison of shape of pair correlation functions calculated at $t = 50$ and averaged over 1000 realisations of the discrete model for varying values of persistence parameter $\varphi = 0.2, 0.4, 0.6, 0.8, 1.0, P = 1$ (top) and for $P = 0.2, 0.4, 0.6, 0.8$, $\varphi = 1$ (bottom). }
	\label{fig:pcf_second}
\end{figure}

Simulations were performed on a $100 \times 100$ lattice, initially populated at CSR and density 0.5 by agents with parameters $P = 1$ and $\varphi = -1, 0, 1$ respectively. For each simulation the PCF was calculated at $t = 20$, at which time the system was judged to have relaxed to its steady state. Results were averaged over $1000$ realisations for each value of $\varphi$. In Fig.~\ref{fig:pcf}(a), we overlay the averaged PCF with the PCF calculated for a single simulation instance, which we also display as a snapshot at $t = 20$. The PCF profiles reveal a striking contrast between persistent and non-persistent agents. In the non-persistent case ($\varphi = 0$), we observe that the PCF remains very close to unity, indicating negligible deviation from CSR as expected. However, in the persistent case ($\varphi = 1$), we observe that the PCF increases significantly above unity at relatively small separation $m < 5$. We also note the presence of a minor dip in the PCF at $m \approx 6$, as highlighted in the inset. With reference to the snapshot shown, this is consistent with visible spontaneous formation of small, localised agent aggregates, as has been previously reported to result from persistence \cite{Soto2014}. Indeed, for higher levels of persistence achieved using the more general two-parameter model, we observe the formation of much larger aggregates (see Appendix \ref{app:aggregation}). Considering the case $\varphi = -1$, we note the converse observation that the PCF decreases significantly below unity for small separation $m < 3$, and increases to reach a slight peak again near $m \approx 4$ (see inset).

In experimentally relevant contexts, it is useful to consider whether this method is robust given limited observation data. The PCFs computed for a single simulation overlaid in Fig.~\ref{fig:pcf}(a) display behaviour at small distances consistent with our observations for the averaged PCF. From this, it is clear that it is possible to distinguish persistent agent behaviour using the PCF, even from a single lattice snapshot. This observation provides strong support for the experimental utility of the PCF as a means of detecting persistence. 

The effect of the persistence parameter $\varphi$ was further investigated by computing the PCF (averaged over 1000 realisations) for simulations with varying values of $\varphi = 0.2, 0.4, 0.6, 0.8, 1.0$, performed up to $t = 20$ with $P = 1$ and agents initially distributed at CSR with density 0.5. The results shown in Fig.~\ref{fig:pcf_second} (top) clearly show that increasing values of $\varphi$ result in roughly proportional increments in the PCF at $m = 1$, corresponding to the probability of agents having an immediate `neighbour'. Similarly, to investigate the effect of the motility parameter $P$ on the PCF, we performed simulations for $P = 0.2, 0.4, 0.6, 0.8, 1.0$, with $\varphi = 1$ up to $t = 50$ and the same initial configuration as previously. Resulting PCFs were averaged over 1000 realisations and shown also in Fig.~\ref{fig:pcf_second} (bottom). These results show that the PCF is independent of $P$, and so the distribution pattern of agents is solely influenced by persistence parameter $\varphi$. In light of this, further work may investigate the PCF as a macroscopic statistic from which information about the microscopic persistence may be extracted.

\subsection{Pair correlation function for a diffusing swarm}\label{sec:pcf_diffusing}

We now consider a non-uniform swarm of agents diffusing away from an initial condition. A $200\times 200$ lattice was used on which agents were initially distributed on $[80, 120]\times [80, 120]$ with unit density. Simulations were performed up to $t = 500$ with parameters $P = 1$ and $\varphi = -1, 0, 1$ respectively. For each value of $\varphi$, the PCF was averaged over 1000 realisations. For clarity, Fig.~\ref{fig:pcf}(b) displays the differences $\Delta f = f(m+1) - f(m)$ against $m$ for the averaged PCF, overlaid with the same result calculated from a single simulation. The original averaged PCF $f$ is shown in the inset, and a snapshot of a single simulation at $t = 500$ is also displayed. As in the case of a uniform swarm, a sharp increase of the PCF for $m < 5$ is characteristic of positive persistence ($\varphi = 1$) and similarly a sharp decrease for negative persistence ($\varphi = -1$). In contrast to the case of a uniform swarm, varying values of $\varphi$ result in differing agent distribution profiles for diffusing swarms and so the overall PCF varies with both $\varphi$ and time. Despite this, the behaviour of the PCF at small $m$ remains characteristic of motion persistence (see Appendix \ref{app:pcf}). These findings suggest that the PCF may be applicable for extracting information about persistence in both uniform and non-uniform swarms of motile agents.  

\section{Conclusion}

We have presented a memory-based formulation for a lattice-based persistent exclusion process in one and two dimensions. In the basic case of a single species of motion-persistent agents, we have shown that such systems are approximately governed in the continuum limit by nonlinear diffusion equations similar to -- but distinct from -- a recent related result \cite{Teomy2019}. We note that key macroscopic differences arise from subtle differences in rules for agent memory in the respective agent-based model formulations. 
In two dimensions, our analysis has been extended to allow for the superposition of a global drift effect as well as interactions between multiple species of persistent walkers, resulting in systems of nonlinear advection-diffusion equations. To our knowledge, equivalent continuum-limit descriptions of the persistent exclusion process incorporating such effects have not been previously presented. We have also explored the utility of the pair correlation function (PCF) as a possible means of inferring the presence of motion persistence from simulated data. Our conclusions indicate that the PCF is a robust statistic from which persistent agent behaviour can be inferred in various settings. 

Possible directions for extension of this work include application of the derived continuum models to problems of biological or other physical relevance, as well as investigation of the usefulness of the PCF for detecting motion persistence in experimental imaging data. Additionally, consideration of quantitative or empirical methods for corrected mean-field approximations \cite{Simpson2011, Markham2013, Teomy2019} could be useful for developing improved continuum models that better account for spatial correlations in both single-species and multispecies distributions of motion-persistent agents. 
Although we have considered only simple exclusion as our model for interactions between different agents, our approach should be able to be extended to a broad class of interactions between agents, in which the occupancies of a finite set of neighbours influence the choice of step direction in competition with the motion persistence
\cite{Penington2011,Johnston2017}.

\acknowledgments This research was supported by an Australian Research Council Discovery Project (DP140100339) and by Australian Mathematical Sciences Institute Vacation Research Scholarships. AC and BDH thank Kerry Landman for helpful discussions.

\appendix

\section{Generalised two-parameter model}\label{app:generalised_model}

The model as presented in the main paper uses a single parameter $\varphi \in [-1, 1]$ to capture the persistence in agent motion. Whilst the use of only a single parameter simplifies analysis, this may be restrictive since the probability of an agent moving orthogonal to its orientation is forced to be $0.5$. We have considered a two-parameter model which allows more control over the probabilistic rule followed by individual agents.

We now introduce two parameters $\alpha, \beta \in [0, 1]$. Under this model, agents attempt moves \emph{in} the direction of orientation with probability $\beta$, \emph{against} the direction of orientation with probability $(1-\alpha)(1-\beta)$, and in either of the choices \emph{orthogonal} to the direction of orientation with probability $\alpha(1-\beta)/2$. Thus, setting $\alpha = 2/3, \beta = 1/4$ we may recover the case of simple exclusion process. From this we may recover the single-parameter model using the relations

\begin{equation}
        \alpha = 1 - \dfrac{1 - \varphi}{3-\varphi}, \qquad \beta = \dfrac{1 + \varphi}{4}.
        \label{eqn:conversion_formulas}
\end{equation}

A similar analysis to that presented for the one-parameter model can be performed for the two-parameter model. We show here results obtained for the model incorporating persistence and global drift, as the governing equations for the case of persistence without drift can be recovered by setting $\lambda = 1$. Employing mean-field arguments, we obtain the approximate discrete master equation %(\ref{eqn:master_2param_drift}). 
given in Table \ref{table:master_2param_drift}. This can be taken to its continuum limit in a similar fashion as demonstrated in the main paper to yield the PDE

\begin{equation}
    \dfrac{\partial u}{\partial t} = \vec{\nabla} \cdot [ \mathcal{D}(u) \vec{\nabla} u] - \vec{\nabla} \cdot (u\vec{v}(u))
\end{equation}

\noindent with diffusivity and velocity field given by 

\begin{align*}
    \mathcal{D}(u) &= D\left(\dfrac{1 + K\lambda}{1 - K\lambda}\right) (1 - 2K\lambda u),\\
    \vec{v}(u) &= \dfrac{2D}{1 - K\lambda} (1-\lambda)(1 - u) \begin{bmatrix} H \\ V \end{bmatrix}, \\
\end{align*}

\noindent where $K$ is given by 

\begin{equation}
    K = -1 + \alpha - \alpha\beta + 2\beta
\end{equation}

\noindent and the constant $D$ is defined by 
\begin{equation}
    D = \lim_{\Delta, \tau \to 0} \dfrac{\Delta^2 P}{4\tau}.
\end{equation}

\noindent For the case of multiple interacting species, we obtain the system of PDEs

\begin{equation}
		\dfrac{\partial u^{(k)}}{\partial t} = \vec{\nabla} \cdot \left[ \mathcal{D}^{(k)}(\vec{u}) \vec{\nabla} u^{(k)} \right] - \vec{\nabla}\cdot(u^{(k)} \vec{v}^{(k)}(\vec{u}))
\end{equation}

\noindent where the diffusivity and velocity fields are 

\begin{align}
	\mathcal{D}^{(k)}(u) &= D^{(k)}\left(\dfrac{1 + K^{(k)}\lambda^{(k)}}{1 - K^{(k)}\lambda^{(k)}}\right)(1 - C^\text{tot}), \\
	\vec{v}^{(k)}(\vec{u}) &= \dfrac{D^{(k)}}{1 - K^{(k)}\lambda^{(k)}}\left\{ 2(1 - \lambda^{(k)})(1-C^\text{tot}) \begin{bmatrix} H \\ V \end{bmatrix}\right\} \\
		&- D^{(k)} \left( \dfrac{1 + K^{(k)}\lambda^{(k)}}{1 - K^{(k)}\lambda^{(k)}}\right)(1- 2K^{(k)}\lambda^{(k)} ) \vec{\nabla} C^\text{tot},
\end{align}
and
\begin{equation}
D^{(k)} = \lim_{\Delta, \tau \to 0} \dfrac{\Delta^2 P^{(k)}}{4\tau}.
\end{equation}

For this general model, comparison of discrete simulation results to continuum approximations yielded similar conclusions to those presented in the main paper.

\begin{table*}
\begin{align*}
	\hspace{-1cm}\begin{split}
		\vec{P}_{n+1}(i, j) &= (1-P)\vec{P}_n(i, j) \\
		&\quad + P\left\{ \left(\lambda \mathrm{diag}\begin{bmatrix} \beta \\ (1-\alpha)(1-\beta) \\ \alpha(1-\beta)/2 \\ \alpha(1-\beta)/2 \end{bmatrix} + (1-\lambda) \dfrac{1+h}{4} \vec{I} \right) C_n(i+1, j)\right.\\
			&\left.\hspace{0.8cm}+ \left(\lambda \mathrm{diag} \begin{bmatrix} (1-\alpha)(1-\beta) \\ \beta \\ \alpha(1-\beta)/2 \\ \alpha(1-\beta)/2 \end{bmatrix} + (1-\lambda) \dfrac{1-h}{4} \vec{I} \right) C_n(i-1, j)\right.\\
			&\left.\hspace{0.8cm}+ \left(\lambda \mathrm{diag} \begin{bmatrix} \alpha(1-\beta)/2 \\ \alpha(1-\beta)/2 \\ \beta \\ (1-\alpha)(1-\beta) \end{bmatrix} + (1-\lambda) \dfrac{1+v}{4} \vec{I} \right) C_n(i, j+1)\right.\\
			&\left.\hspace{0.8cm}+ \left(\lambda \mathrm{diag} \begin{bmatrix} \alpha(1-\beta)/2 \\ \alpha(1-\beta)/2 \\ (1-\alpha)(1-\beta) \\ \beta \end{bmatrix} + (1-\lambda) \dfrac{1-v}{4} \vec{I} \right) C_n(i, j-1)\right\} \vec{P}_n(i, j)\\
		&\quad + P(1-C_n(x, y)) \times \\
		&\qquad\left\{ 
			\left( \begin{bmatrix} \beta & (1-\alpha)(1-\beta) & \alpha(1-\beta)/2 & \alpha(1-\beta)/2 \\ 0&0&0&0 \\ 0&0&0&0 \\ 0&0&0&0 \end{bmatrix} + (1-\lambda) \dfrac{1+h}{4} \begin{bmatrix} 1 & 1 & 1 & 1 \\ 0&0&0&0 \\ 0&0&0&0 \\ 0&0&0&0 \end{bmatrix}\right) \vec{P}_n(i-1, j) \right. \\
				&\qquad\left. + \left( \begin{bmatrix} 0&0&0&0 \\ (1-\alpha)(1-\beta) & \beta & \alpha(1-\beta)/2 & \alpha(1-\beta)/2 \\ 0&0&0&0 \\ 0&0&0&0 \end{bmatrix} + (1-\lambda) \dfrac{1-h}{4} \begin{bmatrix} 0&0&0&0 \\ 1 & 1 & 1 & 1 \\ 0&0&0&0 \\ 0&0&0&0 \end{bmatrix}\right) \vec{P}_n(i+1, j) \right.\\
					&\qquad\left. + \left( \begin{bmatrix} 0&0&0&0 \\ 0&0&0&0 \\ \alpha(1-\beta)/2 & \alpha(1-\beta)/2 & \beta & (1-\alpha)(1-\beta) \\ 0&0&0&0 \end{bmatrix} + (1-\lambda) \dfrac{1+v}{4} \begin{bmatrix} 0&0&0&0 \\ 0&0&0&0 \\ 1 & 1 & 1 & 1 \\ 0&0&0&0 \end{bmatrix}  \right) \vec{P}_n(i, j-1) \right.\\
						&\qquad\left. + \left( \begin{bmatrix} 0&0&0&0 \\ 0&0&0&0 \\ 0&0&0&0 \\ \alpha(1-\beta)/2 & \alpha(1-\beta)/2 & (1-\alpha)(1-\beta) & \beta \end{bmatrix} + (1-\lambda) \dfrac{1-v}{4} \begin{bmatrix} 0&0&0&0 \\ 0&0&0&0 \\ 0&0&0&0 \\ 1 & 1 & 1 & 1 \end{bmatrix} \right) \vec{P}_n(i, j+1)
			\right\}
	\end{split}
%	\label{eqn:master_2param_drift}
\end{align*}
\caption{Master equation for the generalised two-parameter model.}\label{table:master_2param_drift}
\end{table*}

\vfill

\section{Spontaneous aggregation for strong persistence}\label{app:aggregation}

\begin{figure*}
	\centering
	\includegraphics[width = \linewidth]{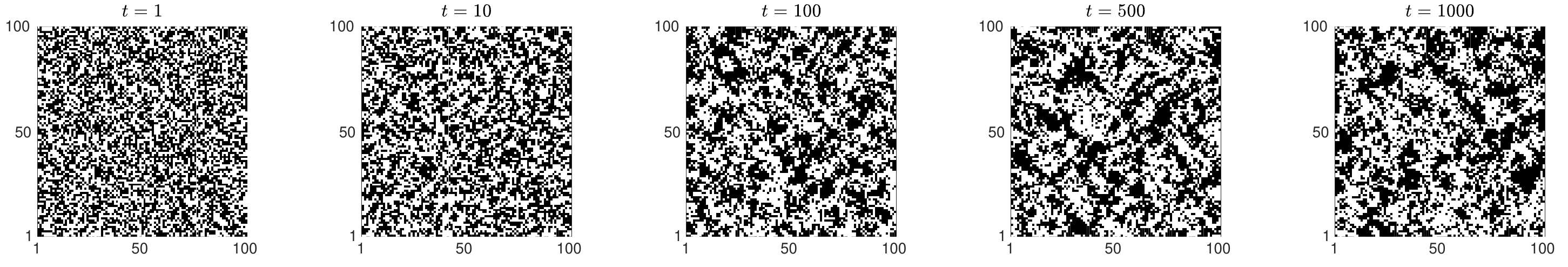}
	\caption{Snapshots of a single simulation for agents with strong persistence ($\beta = 0.9, \alpha = 2/3$) initially distributed at complete spatial randomness with density 0.5. Gradual formation of spontaneous, slowly-evolving aggregates is especially evident for $t \ge 100$.} 
	\label{fig:aggregation}
\end{figure*}

Using the two-parameter model, agents with a strong persistence were modelled by setting $\beta = 0.9, \alpha = 2/3$. Agents were initially distributed at complete spatial randomness at density 0.5, and a single simulation was performed up to $t = 1000$ on a $100\times 100$ square lattice with periodic boundary conditions. Fig.~\ref{fig:aggregation} shows lattice snapshots at several times, revealing clear formation of slowly-evolving aggregates.  

\section{PCF behaviour at small $m$ remains characteristic of persistence}\label{app:pcf}

\begin{figure*}
	\centering
	\includegraphics[width = \linewidth]{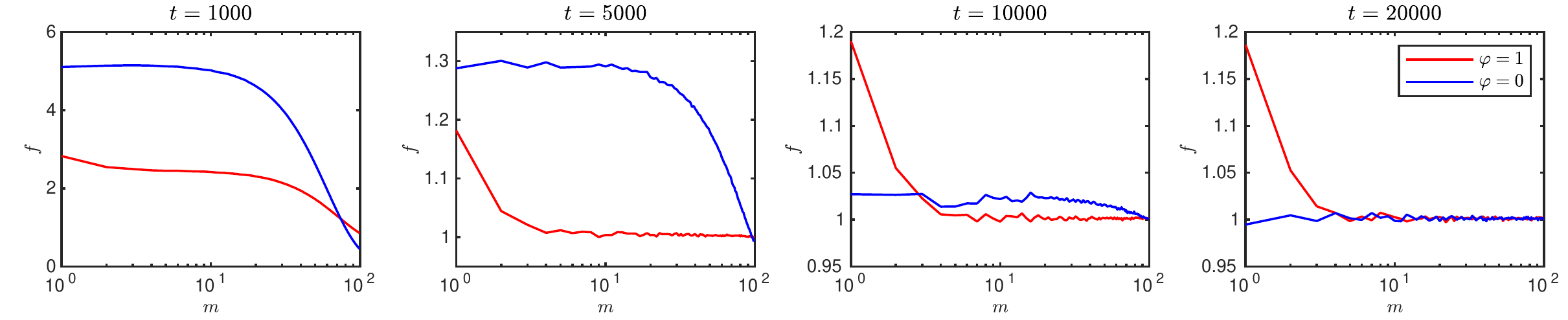}
	\caption{PCFs calculated from simulations of diffusing agent swarms with $\varphi = 0, 1$ at large times, showing the distinguishing behaviour at small $m$ characteristic of persistent motion (compare red and blue curves). The overall change in the PCF profile results from dispersal of bulk population.} 
	\label{fig:large_time}
\end{figure*}

In order to further examine the behaviour of the PCF for a diffusing swarm of persistent agents, simulations were performed as in Section \ref{sec:pcf_diffusing}. in the main paper for $\varphi = 0$ (non-persistent) and $\varphi = 1$ (persistent) agents. Simulations were advanced up to time $t = 20000$, by which time the agents were judged to have fully and uniformly dispersed. PCFs were calculated at $t = 1000, 5000, 10000, 20000$ and averaged over 100 realisations. As shown in Fig.~\ref{fig:large_time}, the characteristic behaviour of the PCF at small values of $m$ as discussed in the main paper is clearly evident across the full timescale of agent dispersal, confirming that it is not obscured by the bulk dispersal of agent population. 

%\pagebreak

%\bibliography{references-with-abbreviated-journal-names}

\begin{thebibliography}{36}%
\makeatletter
\providecommand \@ifxundefined [1]{%
 \@ifx{#1\undefined}
}%
\providecommand \@ifnum [1]{%
 \ifnum #1\expandafter \@firstoftwo
 \else \expandafter \@secondoftwo
 \fi
}%
\providecommand \@ifx [1]{%
 \ifx #1\expandafter \@firstoftwo
 \else \expandafter \@secondoftwo
 \fi
}%
\providecommand \natexlab [1]{#1}%
\providecommand \enquote  [1]{``#1''}%
\providecommand \bibnamefont  [1]{#1}%
\providecommand \bibfnamefont [1]{#1}%
\providecommand \citenamefont [1]{#1}%
\providecommand \href@noop [0]{\@secondoftwo}%
\providecommand \href [0]{\begingroup \@sanitize@url \@href}%
\providecommand \@href[1]{\@@startlink{#1}\@@href}%
\providecommand \@@href[1]{\endgroup#1\@@endlink}%
\providecommand \@sanitize@url [0]{\catcode `\\12\catcode `\$12\catcode
  `\&12\catcode `\#12\catcode `\^12\catcode `\_12\catcode `\%12\relax}%
\providecommand \@@startlink[1]{}%
\providecommand \@@endlink[0]{}%
\providecommand \url  [0]{\begingroup\@sanitize@url \@url }%
\providecommand \@url [1]{\endgroup\@href {#1}{\urlprefix }}%
\providecommand \urlprefix  [0]{URL }%
\providecommand \Eprint [0]{\href }%
\providecommand \doibase [0]{https://doi.org/}%
\providecommand \selectlanguage [0]{\@gobble}%
\providecommand \bibinfo  [0]{\@secondoftwo}%
\providecommand \bibfield  [0]{\@secondoftwo}%
\providecommand \translation [1]{[#1]}%
\providecommand \BibitemOpen [0]{}%
\providecommand \bibitemStop [0]{}%
\providecommand \bibitemNoStop [0]{.\EOS\space}%
\providecommand \EOS [0]{\spacefactor3000\relax}%
\providecommand \BibitemShut  [1]{\csname bibitem#1\endcsname}%
\let\auto@bib@innerbib\@empty
%</preamble>
\bibitem [{\citenamefont {Hughes}(1995)}]{Hughes1995}%
  \BibitemOpen
  \bibfield  {author} {\bibinfo {author} {\bibfnamefont {B.~D.}\ \bibnamefont
  {Hughes}},\ }\href@noop {} {\emph {\bibinfo {title} {Random Walks and Random
  Environments}}},\ Vol.~\bibinfo {volume} {1}\ (\bibinfo  {publisher} {Oxford
  University Press},\ \bibinfo {year} {1995})\BibitemShut {NoStop}%
\bibitem [{\citenamefont {Weiss}(1994)}]{Weiss1994}%
  \BibitemOpen
  \bibfield  {author} {\bibinfo {author} {\bibfnamefont {G.~H.}\ \bibnamefont
  {Weiss}},\ }\href@noop {} {\emph {\bibinfo {title} {Aspects and Applications
  of the Random Walk}}}\ (\bibinfo  {publisher} {North-Holland},\ \bibinfo
  {address} {Amsterdam},\ \bibinfo {year} {1994})\BibitemShut {NoStop}%
\bibitem [{\citenamefont {Codling}\ \emph {et~al.}(2008)\citenamefont
  {Codling}, \citenamefont {Plank},\ and\ \citenamefont
  {Benhamou}}]{Codling2008}%
  \BibitemOpen
  \bibfield  {author} {\bibinfo {author} {\bibfnamefont {E.~A.}\ \bibnamefont
  {Codling}}, \bibinfo {author} {\bibfnamefont {M.~J.}\ \bibnamefont {Plank}},\
  and\ \bibinfo {author} {\bibfnamefont {S.}~\bibnamefont {Benhamou}},\
  }\bibfield  {title} {\bibinfo {title} {{Random walk models in biology}},\
  }\href {https://doi.org/10.1098/rsif.2008.0014} {\bibfield  {journal}
  {\bibinfo  {journal} {J. Roy. Soc. Interface}\ }\textbf {\bibinfo {volume}
  {5}},\ \bibinfo {pages} {813} (\bibinfo {year} {2008})}\BibitemShut {NoStop}%
\bibitem [{\citenamefont {Mort}\ \emph {et~al.}(2016)\citenamefont {Mort},
  \citenamefont {Ross}, \citenamefont {Hainey}, \citenamefont {Harrison},
  \citenamefont {Keighren}, \citenamefont {Landini}, \citenamefont {Baker},
  \citenamefont {Painter}, \citenamefont {Jackson},\ and\ \citenamefont
  {Yates}}]{Mort2016}%
  \BibitemOpen
  \bibfield  {author} {\bibinfo {author} {\bibfnamefont {R.~L.}\ \bibnamefont
  {Mort}}, \bibinfo {author} {\bibfnamefont {R.~J.}\ \bibnamefont {Ross}},
  \bibinfo {author} {\bibfnamefont {K.~J.}\ \bibnamefont {Hainey}}, \bibinfo
  {author} {\bibfnamefont {O.~J.}\ \bibnamefont {Harrison}}, \bibinfo {author}
  {\bibfnamefont {M.~A.}\ \bibnamefont {Keighren}}, \bibinfo {author}
  {\bibfnamefont {G.}~\bibnamefont {Landini}}, \bibinfo {author} {\bibfnamefont
  {R.~E.}\ \bibnamefont {Baker}}, \bibinfo {author} {\bibfnamefont {K.~J.}\
  \bibnamefont {Painter}}, \bibinfo {author} {\bibfnamefont {I.~J.}\
  \bibnamefont {Jackson}},\ and\ \bibinfo {author} {\bibfnamefont {C.~A.}\
  \bibnamefont {Yates}},\ }\bibfield  {title} {\bibinfo {title} {{Reconciling
  diverse mammalian pigmentation patterns with a fundamental mathematical
  model}},\ }\href {https://doi.org/10.1038/ncomms10288} {\bibfield  {journal}
  {\bibinfo  {journal} {Nature Comm.}\ }\textbf {\bibinfo {volume} {7}},\
  \bibinfo {pages} {10288} (\bibinfo {year} {2016})}\BibitemShut {NoStop}%
\bibitem [{\citenamefont {Krummel}\ \emph {et~al.}(2016)\citenamefont
  {Krummel}, \citenamefont {Bartumeus},\ and\ \citenamefont
  {G{\'{e}}rard}}]{Krummel2016}%
  \BibitemOpen
  \bibfield  {author} {\bibinfo {author} {\bibfnamefont {M.~F.}\ \bibnamefont
  {Krummel}}, \bibinfo {author} {\bibfnamefont {F.}~\bibnamefont {Bartumeus}},\
  and\ \bibinfo {author} {\bibfnamefont {A.}~\bibnamefont {G{\'{e}}rard}},\
  }\bibfield  {title} {\bibinfo {title} {{T-cell Migration, Search Strategies
  and Mechanisms}},\ }\href {https://doi.org/10.1038/nri.2015.16.T-cell}
  {\bibfield  {journal} {\bibinfo  {journal} {Nat Rev Immunology}\ }\textbf
  {\bibinfo {volume} {16}},\ \bibinfo {pages} {193} (\bibinfo {year}
  {2016})}\BibitemShut {NoStop}%
\bibitem [{\citenamefont {Baker}\ \emph {et~al.}(2014)\citenamefont {Baker},
  \citenamefont {Yadav}, \citenamefont {Motsch}, \citenamefont {Koschmann},
  \citenamefont {Calinescu}, \citenamefont {Mineharu}, \citenamefont
  {Camelo-Piragua}, \citenamefont {Orringer}, \citenamefont {Bannykh},
  \citenamefont {Nichols}, \citenamefont {DeCarvalho}, \citenamefont
  {Mikkelsen}, \citenamefont {Castro},\ and\ \citenamefont
  {Lowenstein}}]{Baker2014}%
  \BibitemOpen
  \bibfield  {author} {\bibinfo {author} {\bibfnamefont {G.~J.}\ \bibnamefont
  {Baker}}, \bibinfo {author} {\bibfnamefont {V.~N.}\ \bibnamefont {Yadav}},
  \bibinfo {author} {\bibfnamefont {S.}~\bibnamefont {Motsch}}, \bibinfo
  {author} {\bibfnamefont {C.}~\bibnamefont {Koschmann}}, \bibinfo {author}
  {\bibfnamefont {A.~A.}\ \bibnamefont {Calinescu}}, \bibinfo {author}
  {\bibfnamefont {Y.}~\bibnamefont {Mineharu}}, \bibinfo {author}
  {\bibfnamefont {S.~I.}\ \bibnamefont {Camelo-Piragua}}, \bibinfo {author}
  {\bibfnamefont {D.}~\bibnamefont {Orringer}}, \bibinfo {author}
  {\bibfnamefont {S.}~\bibnamefont {Bannykh}}, \bibinfo {author} {\bibfnamefont
  {W.~S.}\ \bibnamefont {Nichols}}, \bibinfo {author} {\bibfnamefont {A.~C.}\
  \bibnamefont {DeCarvalho}}, \bibinfo {author} {\bibfnamefont
  {T.}~\bibnamefont {Mikkelsen}}, \bibinfo {author} {\bibfnamefont {M.~G.}\
  \bibnamefont {Castro}},\ and\ \bibinfo {author} {\bibfnamefont {P.~R.}\
  \bibnamefont {Lowenstein}},\ }\bibfield  {title} {\bibinfo {title}
  {{Mechanisms of glioma formation: Iterative perivascular glioma growth and
  invasion leads to tumor progression, VEGF-independent vascularization, and
  resistance to antiangiogenic therapy}},\ }\href
  {https://doi.org/10.1016/j.neo.2014.06.003} {\bibfield  {journal} {\bibinfo
  {journal} {Neoplasia}\ }\textbf {\bibinfo {volume} {16}},\ \bibinfo {pages}
  {543} (\bibinfo {year} {2014})}\BibitemShut {NoStop}%
\bibitem [{\citenamefont {Simpson}\ \emph
  {et~al.}(2009{\natexlab{a}})\citenamefont {Simpson}, \citenamefont
  {Landman},\ and\ \citenamefont {Hughes}}]{Simpson2009}%
  \BibitemOpen
  \bibfield  {author} {\bibinfo {author} {\bibfnamefont {M.}~\bibnamefont
  {Simpson}}, \bibinfo {author} {\bibfnamefont {K.~A.}\ \bibnamefont
  {Landman}},\ and\ \bibinfo {author} {\bibfnamefont {B.~D.}\ \bibnamefont
  {Hughes}},\ }\bibfield  {title} {\bibinfo {title} {{Diffusing populations:
  ghosts or folks}},\ }\href {https://doi.org/10.1080/22054952.2009.11464027}
  {\bibfield  {journal} {\bibinfo  {journal} {Australas. J. Eng. Edu.}\
  }\textbf {\bibinfo {volume} {15}},\ \bibinfo {pages} {93} (\bibinfo {year}
  {2009}{\natexlab{a}})}\BibitemShut {NoStop}%
\bibitem [{\citenamefont {Liggett}(1985)}]{Liggett1985}%
  \BibitemOpen
  \bibfield  {author} {\bibinfo {author} {\bibfnamefont {T.~M.}\ \bibnamefont
  {Liggett}},\ }\href@noop {} {\emph {\bibinfo {title} {Interacting Particle
  Systems}}}\ (\bibinfo  {publisher} {Springer-Verlag},\ \bibinfo {address}
  {New York},\ \bibinfo {year} {1985})\BibitemShut {NoStop}%
\bibitem [{\citenamefont {Liggett}(1999)}]{Liggett1999}%
  \BibitemOpen
  \bibfield  {author} {\bibinfo {author} {\bibfnamefont {T.~M.}\ \bibnamefont
  {Liggett}},\ }\href@noop {} {\emph {\bibinfo {title} {Stochastic Interacting
  Systems: Contact, Voter and Exclusion Processes}}}\ (\bibinfo  {publisher}
  {Springer-Verlag},\ \bibinfo {address} {Berlin},\ \bibinfo {year}
  {1999})\BibitemShut {NoStop}%
\bibitem [{\citenamefont {Komorowski}\ \emph {et~al.}(2007)\citenamefont
  {Komorowski}, \citenamefont {Landim},\ and\ \citenamefont
  {Olla}}]{Komorowski2007}%
  \BibitemOpen
  \bibfield  {author} {\bibinfo {author} {\bibfnamefont {T.}~\bibnamefont
  {Komorowski}}, \bibinfo {author} {\bibfnamefont {C.}~\bibnamefont {Landim}},\
  and\ \bibinfo {author} {\bibfnamefont {S.}~\bibnamefont {Olla}},\ }\href
  {https://doi.org/10.1007/978-3-642-29880-6} {\emph {\bibinfo {title}
  {{Fluctuations in Markov Processes}}}}\ (\bibinfo  {publisher} {Springer},\
  \bibinfo {address} {Heidelberg},\ \bibinfo {year} {2007})\BibitemShut
  {NoStop}%
\bibitem [{\citenamefont {Markham}\ \emph {et~al.}(2013)\citenamefont
  {Markham}, \citenamefont {Simpson},\ and\ \citenamefont
  {Baker}}]{Markham2013}%
  \BibitemOpen
  \bibfield  {author} {\bibinfo {author} {\bibfnamefont {D.~C.}\ \bibnamefont
  {Markham}}, \bibinfo {author} {\bibfnamefont {M.~J.}\ \bibnamefont
  {Simpson}},\ and\ \bibinfo {author} {\bibfnamefont {R.~E.}\ \bibnamefont
  {Baker}},\ }\bibfield  {title} {\bibinfo {title} {{Simplified method for
  including spatial correlations in mean-field approximations}},\ }\bibfield
  {journal} {\bibinfo  {journal} {Phys. Rev. E}\ }\textbf {\bibinfo {volume}
  {87}},\ \bibinfo {pages} {062702} (\bibinfo {year} {2013})\BibitemShut {NoStop}%
\bibitem [{\citenamefont {Simpson}\ \emph
  {et~al.}(2009{\natexlab{b}})\citenamefont {Simpson}, \citenamefont
  {Landman},\ and\ \citenamefont {Hughes}}]{Simpson2009b}%
  \BibitemOpen
  \bibfield  {author} {\bibinfo {author} {\bibfnamefont {M.~J.}\ \bibnamefont
  {Simpson}}, \bibinfo {author} {\bibfnamefont {K.~A.}\ \bibnamefont
  {Landman}},\ and\ \bibinfo {author} {\bibfnamefont {B.~D.}\ \bibnamefont
  {Hughes}},\ }\bibfield  {title} {\bibinfo {title} {{Multi-species simple
  exclusion processes}},\ }\href {https://doi.org/10.1016/j.physa.2008.10.038}
  {\bibfield  {journal} {\bibinfo  {journal} {Physica A}\ }\textbf {\bibinfo
  {volume} {388}},\ \bibinfo {pages} {399} (\bibinfo {year}
  {2009}{\natexlab{b}})}\BibitemShut {NoStop}%
\bibitem [{\citenamefont {Penington}\ \emph {et~al.}(2011)\citenamefont
  {Penington}, \citenamefont {Hughes},\ and\ \citenamefont
  {Landman}}]{Penington2011}%
  \BibitemOpen
  \bibfield  {author} {\bibinfo {author} {\bibfnamefont {C.~J.}\ \bibnamefont
  {Penington}}, \bibinfo {author} {\bibfnamefont {B.~D.}\ \bibnamefont
  {Hughes}},\ and\ \bibinfo {author} {\bibfnamefont {K.~A.}\ \bibnamefont
  {Landman}},\ }\bibfield  {title} {\bibinfo {title} {{Building macroscale
  models from microscale probabilistic models: A general probabilistic approach
  for nonlinear diffusion and multispecies phenomena}},\ }\href
  {https://doi.org/10.1103/PhysRevE.84.041120} {\bibfield  {journal} {\bibinfo
  {journal} {Phys. Rev. E}\ }\textbf {\bibinfo {volume} {84}},\ \bibinfo
  {pages} {041120} (\bibinfo {year} {2011})}\BibitemShut {NoStop}%
\bibitem [{\citenamefont {Almet}\ \emph {et~al.}(2015)\citenamefont {Almet},
  \citenamefont {Pan}, \citenamefont {Hughes},\ and\ \citenamefont
  {Landman}}]{Almet2015}%
  \BibitemOpen
  \bibfield  {author} {\bibinfo {author} {\bibfnamefont {A.~A.}\ \bibnamefont
  {Almet}}, \bibinfo {author} {\bibfnamefont {M.}~\bibnamefont {Pan}}, \bibinfo
  {author} {\bibfnamefont {B.~D.}\ \bibnamefont {Hughes}},\ and\ \bibinfo
  {author} {\bibfnamefont {K.~A.}\ \bibnamefont {Landman}},\ }\bibfield
  {title} {\bibinfo {title} {{When push comes to shove: Exclusion processes
  with nonlocal consequences}},\ }\href
  {https://doi.org/10.1016/j.physa.2015.05.031} {\bibfield  {journal} {\bibinfo
   {journal} {Physica A}\ }\textbf {\bibinfo {volume} {437}},\ \bibinfo {pages}
  {119} (\bibinfo {year} {2015})}\BibitemShut {NoStop}%
\bibitem [{\citenamefont {Nan}\ \emph {et~al.}(2018)\citenamefont {Nan},
  \citenamefont {Walsh}, \citenamefont {Landman},\ and\ \citenamefont
  {Hughes}}]{Nan2018}%
  \BibitemOpen
  \bibfield  {author} {\bibinfo {author} {\bibfnamefont {P.}~\bibnamefont
  {Nan}}, \bibinfo {author} {\bibfnamefont {D.~M.}\ \bibnamefont {Walsh}},
  \bibinfo {author} {\bibfnamefont {K.~A.}\ \bibnamefont {Landman}},\ and\
  \bibinfo {author} {\bibfnamefont {B.~D.}\ \bibnamefont {Hughes}},\ }\bibfield
   {title} {\bibinfo {title} {{Distinguishing cell shoving mechanisms}},\
  }\href {https://doi.org/10.1371/journal.pone.0193975} {\bibfield  {journal}
  {\bibinfo  {journal} {PLoS ONE}\ }\textbf {\bibinfo {volume} {13}},\ \bibinfo
  {pages} {e0193975} (\bibinfo {year} {2018})}\BibitemShut {NoStop}%
\bibitem [{\citenamefont {Fernando}\ \emph {et~al.}(2010)\citenamefont
  {Fernando}, \citenamefont {Landman},\ and\ \citenamefont
  {Simpson}}]{Fernando2010}%
  \BibitemOpen
  \bibfield  {author} {\bibinfo {author} {\bibfnamefont {A.~E.}\ \bibnamefont
  {Fernando}}, \bibinfo {author} {\bibfnamefont {K.~A.}\ \bibnamefont
  {Landman}},\ and\ \bibinfo {author} {\bibfnamefont {M.~J.}\ \bibnamefont
  {Simpson}},\ }\bibfield  {title} {\bibinfo {title} {{Nonlinear diffusion and
  exclusion processes with contact interactions}},\ }\href
  {https://doi.org/10.1103/PhysRevE.81.011903} {\bibfield  {journal} {\bibinfo
  {journal} {Phys. Rev. E}\ }\textbf {\bibinfo {volume} {81}},\ \bibinfo
  {pages} {011903} (\bibinfo {year} {2010})}\BibitemShut {NoStop}%
\bibitem [{\citenamefont {Michelini}\ and\ \citenamefont
  {Coyle}(2008)}]{Michelini2008}%
  \BibitemOpen
  \bibfield  {author} {\bibinfo {author} {\bibfnamefont {P.~N.}\ \bibnamefont
  {Michelini}}\ and\ \bibinfo {author} {\bibfnamefont {E.~J.}\ \bibnamefont
  {Coyle}},\ }\bibfield  {title} {\bibinfo {title} {Mobility models based on
  correlated random walks},\ }in\ {\emph {\bibinfo {booktitle}
  {Proceedings of the International Conference on Mobile Technology,
  Applications, and Systems}}},\ \bibinfo {series and number} {Mobility '08}\
  (\bibinfo  {publisher} {ACM},\ \bibinfo {address} {New York, NY, USA},\
  \bibinfo {year} {2008})\ pp.\ \bibinfo {pages} {86:1--86:8}, \href
  {https://doi.org/10.1145/1506270.1506376} {https://doi.org/10.1145/1506270.1506376}\BibitemShut
  {NoStop}%
\bibitem [{\citenamefont {Vilensky}\ \emph {et~al.}(1994)\citenamefont
  {Vilensky}, \citenamefont {Havlin}, \citenamefont {Taitelbaum},\ and\
  \citenamefont {Weiss}}]{Vilensky1994}%
  \BibitemOpen
  \bibfield  {author} {\bibinfo {author} {\bibfnamefont {B.}~\bibnamefont
  {Vilensky}}, \bibinfo {author} {\bibfnamefont {S.}~\bibnamefont {Havlin}},
  \bibinfo {author} {\bibfnamefont {H.}~\bibnamefont {Taitelbaum}},\ and\
  \bibinfo {author} {\bibfnamefont {G.~H.}\ \bibnamefont {Weiss}},\ }\bibfield
  {title} {\bibinfo {title} {Momentum effects in reaction-diffusion systems},\
  }\href@noop {} {\bibfield  {journal} {\bibinfo  {journal} {J. Phys. Chem.}\
  }\textbf {\bibinfo {volume} {98}},\ \bibinfo {pages} {7325} (\bibinfo {year}
  {1994})}\BibitemShut {NoStop}%
\bibitem [{\citenamefont {Wu}\ \emph {et~al.}(2000)\citenamefont {Wu},
  \citenamefont {Li}, \citenamefont {Springer},\ and\ \citenamefont
  {Neill}}]{Wu2000}%
  \BibitemOpen
  \bibfield  {author} {\bibinfo {author} {\bibfnamefont {H.~I.}\ \bibnamefont
  {Wu}}, \bibinfo {author} {\bibfnamefont {B.~L.}\ \bibnamefont {Li}}, \bibinfo
  {author} {\bibfnamefont {T.~A.}\ \bibnamefont {Springer}},\ and\ \bibinfo
  {author} {\bibfnamefont {W.~H.}\ \bibnamefont {Neill}},\ }\bibfield  {title}
  {\bibinfo {title} {{Modelling animal movement as a persistent random walk in
  two dimensions: Expected magnitude of net displacement}},\ }\href
  {https://doi.org/10.1016/S0304-3800(00)00309-4} {\bibfield  {journal}
  {\bibinfo  {journal} {Ecological Modelling}\ }\textbf {\bibinfo {volume}
  {132}},\ \bibinfo {pages} {115} (\bibinfo {year} {2000})}\BibitemShut
  {NoStop}%
\bibitem [{\citenamefont {Gorelik}\ and\ \citenamefont
  {Gautreau}(2014)}]{Gorelik2014}%
  \BibitemOpen
  \bibfield  {author} {\bibinfo {author} {\bibfnamefont {R.}~\bibnamefont
  {Gorelik}}\ and\ \bibinfo {author} {\bibfnamefont {A.}~\bibnamefont
  {Gautreau}},\ }\bibfield  {title} {\bibinfo {title} {{Quantitative and
  unbiased analysis of directional persistence in cell migration}},\ }\href
  {https://doi.org/10.1038/nprot.2014.131} {\bibfield  {journal} {\bibinfo
  {journal} {Nature Protocols}\ }\textbf {\bibinfo {volume} {9}},\ \bibinfo
  {pages} {1931} (\bibinfo {year} {2014})}\BibitemShut {NoStop}%
\bibitem [{\citenamefont {Patlak}(1953)}]{Patlak1953}%
  \BibitemOpen
  \bibfield  {author} {\bibinfo {author} {\bibfnamefont {C.~S.}\ \bibnamefont
  {Patlak}},\ }\bibfield  {title} {\bibinfo {title} {{Random walk with
  persistence and external bias}},\ }\href {https://doi.org/10.1007/BF02476407}
  {\bibfield  {journal} {\bibinfo  {journal} {Bull. Math. Biophys.}\ }\textbf
  {\bibinfo {volume} {15}},\ \bibinfo {pages} {311} (\bibinfo {year}
  {1953})}\BibitemShut {NoStop}%
\bibitem [{\citenamefont {Masoliver}\ and\ \citenamefont
  {Lindenberg}(2017)}]{Masoliver2017}%
  \BibitemOpen
  \bibfield  {author} {\bibinfo {author} {\bibfnamefont {J.}~\bibnamefont
  {Masoliver}}\ and\ \bibinfo {author} {\bibfnamefont {K.}~\bibnamefont
  {Lindenberg}},\ }\bibfield  {title} {\bibinfo {title} {{Continuous time
  persistent random walk: a review and some generalizations}},\ }\href
  {https://doi.org/10.1140/epjb/e2017-80123-7} {\bibfield  {journal} {\bibinfo
  {journal} {Eur. Phys. J. B}\ }\textbf {\bibinfo {volume} {90}},\ \bibinfo
  {pages} {107} (\bibinfo {year} {2017})}\BibitemShut {NoStop}%
\bibitem [{\citenamefont {Renshaw}\ and\ \citenamefont
  {Henderson}(1981)}]{Renshaw1981}%
  \BibitemOpen
  \bibfield  {author} {\bibinfo {author} {\bibfnamefont {E.}~\bibnamefont
  {Renshaw}}\ and\ \bibinfo {author} {\bibfnamefont {R.}~\bibnamefont
  {Henderson}},\ }\bibfield  {title} {\bibinfo {title} {{The correlated random
  walk}},\ }\href {https://doi.org/10.2307/3213286} {\bibfield  {journal}
  {\bibinfo  {journal} {J. Appl. Probab.}\ }\textbf {\bibinfo {volume} {18}},\
  \bibinfo {pages} {403} (\bibinfo {year} {1981})}\BibitemShut {NoStop}%
\bibitem [{\citenamefont {Henderson}\ \emph {et~al.}(1984)\citenamefont
  {Henderson}, \citenamefont {Renshaw},\ and\ \citenamefont
  {Ford}}]{Henderson1984}%
  \BibitemOpen
  \bibfield  {author} {\bibinfo {author} {\bibfnamefont {R.}~\bibnamefont
  {Henderson}}, \bibinfo {author} {\bibfnamefont {E.}~\bibnamefont {Renshaw}},\
  and\ \bibinfo {author} {\bibfnamefont {D.}~\bibnamefont {Ford}},\ }\bibfield
  {title} {\bibinfo {title} {{Correlated Random Walk Model for Two-Dimensional
  Diffusion.}},\ }\href {https://doi.org/10.2307/3213636} {\bibfield  {journal}
  {\bibinfo  {journal} {J. Appl. Probab.}\ }\textbf {\bibinfo {volume} {21}},\
  \bibinfo {pages} {233} (\bibinfo {year} {1984})}\BibitemShut {NoStop}%
\bibitem [{\citenamefont {Gavagnin}\ and\ \citenamefont
  {Yates}(2018)}]{Gavagnin2018}%
  \BibitemOpen
  \bibfield  {author} {\bibinfo {author} {\bibfnamefont {E.}~\bibnamefont
  {Gavagnin}}\ and\ \bibinfo {author} {\bibfnamefont {C.~A.}\ \bibnamefont
  {Yates}},\ }\bibfield  {title} {\bibinfo {title} {{Modeling persistence of
  motion in a crowded environment: The diffusive limit of excluding
  velocity-jump processes}},\ }\bibfield  {journal} {\bibinfo  {journal} {Phys.
  Rev. E}\ }\textbf {\bibinfo {volume} {97}},\
  \bibinfo {pages} {032416} 
  (\bibinfo {year} {2018}), \Eprint
  {https://arxiv.org/abs/1710.09264v2} {arXiv:1710.09264v2} \BibitemShut
  {NoStop}%
\bibitem [{\citenamefont {Teomy}\ and\ \citenamefont
  {Metzler}(2019{\natexlab{a}})}]{Teomy2019}%
  \BibitemOpen
  \bibfield  {author} {\bibinfo {author} {\bibfnamefont {E.}~\bibnamefont
  {Teomy}}\ and\ \bibinfo {author} {\bibfnamefont {R.}~\bibnamefont
  {Metzler}},\ }\bibfield  {title} {\bibinfo {title} {{Correlations and
  transport in exclusion processes with general finite memory}},\ }\Eprint
  {https://arxiv.org/abs/1906.10447} {arXiv:1906.10447}  (\bibinfo {year}
  {2019}{\natexlab{a}})\BibitemShut {NoStop}%
\bibitem [{\citenamefont {Teomy}\ and\ \citenamefont
  {Metzler}(2019{\natexlab{b}})}]{Teomy2019b}%
  \BibitemOpen
  \bibfield  {author} {\bibinfo {author} {\bibfnamefont {E.}~\bibnamefont
  {Teomy}}\ and\ \bibinfo {author} {\bibfnamefont {R.}~\bibnamefont
  {Metzler}},\ }\bibfield  {title} {\bibinfo {title} {{Transport in exclusion
  processes with one-step memory: density dependence and optimal
  acceleration}},\ }\Eprint {https://arxiv.org/abs/1906.10442}
  {arXiv:1906.10442}  (\bibinfo {year} {2019}{\natexlab{b}})\BibitemShut
  {NoStop}%
\bibitem [{\citenamefont {Jones}\ \emph {et~al.}(2015)\citenamefont {Jones},
  \citenamefont {Sim}, \citenamefont {Taylor}, \citenamefont {Bugeon},
  \citenamefont {Dallman}, \citenamefont {Pereira}, \citenamefont {Stumpf},\
  and\ \citenamefont {Liepe}}]{Jones2015}%
  \BibitemOpen
  \bibfield  {author} {\bibinfo {author} {\bibfnamefont {P.~J.}\ \bibnamefont
  {Jones}}, \bibinfo {author} {\bibfnamefont {A.}~\bibnamefont {Sim}}, \bibinfo
  {author} {\bibfnamefont {H.~B.}\ \bibnamefont {Taylor}}, \bibinfo {author}
  {\bibfnamefont {L.}~\bibnamefont {Bugeon}}, \bibinfo {author} {\bibfnamefont
  {M.~J.}\ \bibnamefont {Dallman}}, \bibinfo {author} {\bibfnamefont
  {B.}~\bibnamefont {Pereira}}, \bibinfo {author} {\bibfnamefont {M.~P.}\
  \bibnamefont {Stumpf}},\ and\ \bibinfo {author} {\bibfnamefont
  {J.}~\bibnamefont {Liepe}},\ }\bibfield  {title} {\bibinfo {title}
  {{Inference of random walk models to describe leukocyte migration}},\
  }\href {https://doi.org/10.1088/1478-3975/12/6/066001}
  {\bibfield  {journal} {\bibinfo  {journal} {Phys. Biol.}\ }\textbf {\bibinfo
  {volume} {12}},\ \bibinfo {pages} {066001}  (\bibinfo {year} {2015})}\BibitemShut {NoStop}%
\bibitem [{\citenamefont {Galanti}\ \emph {et~al.}(2013)\citenamefont
  {Galanti}, \citenamefont {Fanelli},\ and\ \citenamefont
  {Piazza}}]{Galanti2013}%
  \BibitemOpen
  \bibfield  {author} {\bibinfo {author} {\bibfnamefont {M.}~\bibnamefont
  {Galanti}}, \bibinfo {author} {\bibfnamefont {D.}~\bibnamefont {Fanelli}},\
  and\ \bibinfo {author} {\bibfnamefont {F.}~\bibnamefont {Piazza}},\
  }\bibfield  {title} {\bibinfo {title} {{Persistent random walk with
  exclusion}},\ }\href {https://doi.org/10.1140/epjb/e2013-40838-y} {\bibfield
  {journal} {\bibinfo  {journal} {Eur. Phys. J. B}\ }\textbf {\bibinfo {volume}
  {86}},\ \bibinfo {pages} {456} (\bibinfo {year} {2013})},\ \Eprint
  {https://arxiv.org/abs/arXiv:1306.3400v2} {arXiv:1306.3400v2} \BibitemShut
  {NoStop}%
\bibitem [{\citenamefont {Agnew}\ \emph {et~al.}(2014)\citenamefont {Agnew},
  \citenamefont {Green}, \citenamefont {Brown}, \citenamefont {Simpson},\ and\
  \citenamefont {Binder}}]{Agnew2014}%
  \BibitemOpen
  \bibfield  {author} {\bibinfo {author} {\bibfnamefont {D.~J.}\ \bibnamefont
  {Agnew}}, \bibinfo {author} {\bibfnamefont {J.~E.}\ \bibnamefont {Green}},
  \bibinfo {author} {\bibfnamefont {T.~M.}\ \bibnamefont {Brown}}, \bibinfo
  {author} {\bibfnamefont {M.~J.}\ \bibnamefont {Simpson}},\ and\ \bibinfo
  {author} {\bibfnamefont {B.~J.}\ \bibnamefont {Binder}},\ }\bibfield  {title}
  {\bibinfo {title} {{Distinguishing between mechanisms of cell aggregation
  using pair-correlation functions}},\ }\href
  {https://doi.org/10.1016/j.jtbi.2014.02.033} {\bibfield  {journal} {\bibinfo
  {journal} {J. Theoret. Biol.}\ }\textbf {\bibinfo {volume} {352}},\ \bibinfo
  {pages} {16} (\bibinfo {year} {2014})}\BibitemShut {NoStop}%
\bibitem [{\citenamefont {Treloar}\ \emph {et~al.}(2011)\citenamefont
  {Treloar}, \citenamefont {Simpson},\ and\ \citenamefont
  {McCue}}]{Treloar2011}%
  \BibitemOpen
  \bibfield  {author} {\bibinfo {author} {\bibfnamefont {K.~K.}\ \bibnamefont
  {Treloar}}, \bibinfo {author} {\bibfnamefont {M.~J.}\ \bibnamefont
  {Simpson}},\ and\ \bibinfo {author} {\bibfnamefont {S.~W.}\ \bibnamefont
  {McCue}},\ }\bibfield  {title} {\bibinfo {title} {{Velocity-jump models with
  crowding effects}},\ }  \href{https://doi.org/10.1103/PhysRevE.84.061920} {\bibfield  {journal} {\bibinfo  {journal} {Phys. Rev.
  E}\ }\textbf {\bibinfo {volume} {84}},\ \bibinfo {pages} {061920} 
  (\bibinfo {year} {2011})}\BibitemShut {NoStop}%
\bibitem [{\citenamefont {Simpson}\ and\ \citenamefont
  {Baker}(2011)}]{Simpson2011}%
  \BibitemOpen
  \bibfield  {author} {\bibinfo {author} {\bibfnamefont {M.~J.}\ \bibnamefont
  {Simpson}}\ and\ \bibinfo {author} {\bibfnamefont {R.~E.}\ \bibnamefont
  {Baker}},\ }\bibfield  {title} {\bibinfo {title} {{Corrected mean-field
  models for spatially dependent advection-diffusion- reaction phenomena}},\ }  \href{https://doi.org/10.1103/PhysRevE.83.051922}
  {\bibfield  {journal} {\bibinfo  {journal} {Phys. Rev. E}\ }\textbf {\bibinfo {volume} {83}},\
  \bibinfo {pages} {051922} (\bibinfo {year} {2011})}\BibitemShut {NoStop}%
\bibitem [{\citenamefont {Johnston}\ and\ \citenamefont
  {Crampin}(2019)}]{Johnston2019}%
  \BibitemOpen
  \bibfield  {author} {\bibinfo {author} {\bibfnamefont {S.~T.}\ \bibnamefont
  {Johnston}}\ and\ \bibinfo {author} {\bibfnamefont {E.~J.}\ \bibnamefont
  {Crampin}},\ }\bibfield  {title} {\bibinfo {title} {{Corrected pair
  correlation functions for environments with obstacles}},\ } 
   \href{https://doi.org/10.1103/PhysRevE.99.032124}{\bibfield  {journal} {\bibinfo {journal} {Phys. Rev. E}\ }\textbf {\bibinfo {volume} {99}},\ 
  \bibinfo {pages} {032124} (\bibinfo {year} {2019})},\ \Eprint
  {https://arxiv.org/abs/arXiv:1811.07518v1}  {arXiv:1811.07518v1}\BibitemShut {NoStop}%
\bibitem [{\citenamefont {Gavagnin}\ \emph {et~al.}(2018)\citenamefont
  {Gavagnin}, \citenamefont {Owen},\ and\ \citenamefont
  {Yates}}]{Gavagnin2018b}%
  \BibitemOpen
  \bibfield  {author} {\bibinfo {author} {\bibfnamefont {E.}~\bibnamefont
  {Gavagnin}}, \bibinfo {author} {\bibfnamefont {J.~P.}\ \bibnamefont {Owen}},\
  and\ \bibinfo {author} {\bibfnamefont {C.~A.}\ \bibnamefont {Yates}},\
  }\bibfield  {title} {\bibinfo {title} {{Pair correlation functions for
  identifying spatial correlation in discrete domains}},\ }  \href{https://doi.org/10.1103/PhysRevE.97.062104} {\bibfield  {journal}
  {\bibinfo  {journal} {Phys. Rev. E}\ }\textbf {\bibinfo {volume} {97}},\ 
  \bibinfo {pages} {062104} (\bibinfo {year} {2018})},\ \Eprint {https://arxiv.org/abs/1804.03452v2} {arXiv:1804.03452v2}
  \BibitemShut {NoStop}%
\bibitem [{\citenamefont {Deroulers}\ \emph {et~al.}(2009)\citenamefont
  {Deroulers}, \citenamefont {Aubert}, \citenamefont {Badoual},\ and\
  \citenamefont {Grammaticos}}]{Deroulers2009}%
  \BibitemOpen
  \bibfield  {author} {\bibinfo {author} {\bibfnamefont {C.}~\bibnamefont
  {Deroulers}}, \bibinfo {author} {\bibfnamefont {M.}~\bibnamefont {Aubert}},
  \bibinfo {author} {\bibfnamefont {M.}~\bibnamefont {Badoual}},\ and\ \bibinfo
  {author} {\bibfnamefont {B.}~\bibnamefont {Grammaticos}},\ }\bibfield
  {title} {\bibinfo {title} {{Modeling tumor cell migration: From microscopic
  to macroscopic models}},\ }\href {https://doi.org/10.1103/PhysRevE.79.031917}
  {\bibfield  {journal} {\bibinfo  {journal} {Phys. Rev. E}\ }\textbf {\bibinfo
  {volume} {79}},\ \bibinfo {pages} {031917} (\bibinfo {year} {2009})}\BibitemShut
  {NoStop}%
\bibitem [{\citenamefont {Soto}\ and\ \citenamefont
  {Golestanian}(2014)}]{Soto2014}%
  \BibitemOpen
  \bibfield  {author} {\bibinfo {author} {\bibfnamefont {R.}~\bibnamefont
  {Soto}}\ and\ \bibinfo {author} {\bibfnamefont {R.}~\bibnamefont
  {Golestanian}},\ }\bibfield  {title} {\bibinfo {title} {{Run-and-tumble
  dynamics in a crowded environment: Persistent exclusion process for
  swimmers}},\ }\href {https://doi.org/10.1103/PhysRevE.89.012706} {\bibfield  {journal} {\bibinfo  {journal} {Phys. Rev. E}\
  }\textbf {\bibinfo {volume} {89}},\ \bibinfo {pages} {012706}
  (\bibinfo {year} {2014})},\ \Eprint
  {https://arxiv.org/abs/1306.0481v2} {arXiv:1306.0481v2} \BibitemShut
  {NoStop}%
\bibitem [{\citenamefont {Johnston}\ \emph{et~al.}(2017)\citenamefont {Johnston},
  \citenamefont {Baker}, \citenamefont {McElwain}, \ and\ \citenamefont {Simpson}}]{Johnston2017}%%
   \BibitemOpen
  \bibfield {author} {\bibinfo {author} {\bibfnamefont {S.~T.}\ \bibnamefont
  {Johnston}}, \bibinfo {author} {\bibfnamefont {R.~E.}\ \bibnamefont
  {Baker}}, \bibinfo {author} {\bibfnamefont {D.~L.~S.}\ \bibfnamefont {McElwain}}, \bibinfo {author} {\bibfnamefont {M.~J.}\ \bibfnamefont {Simpson}}\ }\bibfield  {title} {\bibinfo {title} {{Co-operation, competition and crowding: a discrete framework linking allee kinetics, nonlinear diffusion, shocks and sharp-fronted travelling waves}},\ } 
   \href{https://doi.org/10.1038/srep42134}{\bibfield  {journal} {\bibinfo {journal} {Sci. Rep.}\ }\textbf {\bibinfo {volume} {7}},\ 
  \bibinfo {pages} {42134} (\bibinfo {year} {2017})}
\BibitemShut {NoStop}%
\end{thebibliography}

\end{document}